\documentclass{aa}     % LaTeX A&A New Fonts
\usepackage{epsfig}
\begin{document}
\def\h50{h$_{50}^{-1}${}}
\def\kms{km~s$^{-1}${}}
\thesaurus{11.01.2; 11.06.2; 11.09.2; 11.16.1; 11.17.3; 13.09.1}

\title{Near infrared observations of quasars with extended ionized
envelopes \thanks{Based on data obtained at the European Southern
Observatory, La Silla, Chile.  Also based on observations made with
the NASA/ESA Hubble Space Telescope, obtained from the data archive at
the Space Telescope Science Institute; STScI is operated by the
Association of Universities for Research in Astronomy, Inc. under NASA
contract NAS 5-26555. This research has made use of the SIMBAD
database, operated at CDS, Strasbourg, France and of the NASA/IPAC
extragalactic database (NED), which is operated by the Jet Propulsion
Laboratory, Caltech under contract with the National Aeronautics and
Space Administration.}}

\author{
  I.~M\'arquez \inst{1,2}
\and
  F.~Durret \inst{2,3}
\and
  P.~Petitjean \inst{2,3}  
}
\offprints{I. M\'arquez (\sl{isabel@iaa.es}) }
\institute{
	Instituto de Astrof\'\i sica de Andaluc\'\i a (C.S.I.C.), 
Apartado 3004 , E-18080 Granada, Spain
\and
	Institut d'Astrophysique de Paris, CNRS, 98bis Bd Arago, 
F-75014 Paris, France 
\and 
	DAEC, Observatoire de Paris, Universit\'e Paris VII, CNRS (UA 173), 
F-92195 Meudon Cedex, France 
}
\date{Received,  ; accepted,}

\maketitle
\markboth{M\'arquez et al.: Underlying galaxies in a sample of 
quasars with extended ionized envelopes.}{}

\begin{abstract}

We have observed a sample of 15 and 8 quasars with redshifts between
0.11 and 0.87 (mean value 0.38) in the J and K' bands
respectively. Eleven of the quasars were previously known to be
associated with extended emission line regions.  After deconvolution
of the image, substraction of the PSF when possible, and
identification of companions with the help of HST archive images when
available, extensions are seen for at least eleven quasars.  However,
average profiles are different from that of the PSF in only four objects,
for which a good fit is obtained with an $r^{1/4}$ law, suggesting
that the underlying galaxies are ellipticals.

Redshifts were available in the literature for surrounding objects in
five quasar fields. For these objects, one to five companion galaxies
were found. One quasar even belongs to a richness class 1
cluster. Most other quasars in our sample have nearby galaxies in
projection which may also be companions. Environmental effects are
therefore probably important to account for the properties of these
objects.

\end{abstract}

\section{Introduction}

Attempts to understand properties of active galactic nuclei (AGN) and
their cosmological evolution has led to the conclusion that the AGN
activity as a whole is most likely to be described as a succession of
episodic and time limited events happening in most if not all galaxies
(Cavaliere \& Padovani 1989, Collin-Souffrin 1991).  The most
plausible explanation for the onset of activity is fuelling of the
nucleus by gas streaming towards the center of the galaxy as a result
of interaction, accretion or merging events.

AGN surrounded by an extended emission line region are of particular
interest since most often gas lying very far from the center is
revealed by the ionizing flux from the active nucleus.  Studying the
kinematics and physical properties of this gas is a unique way towards
disentangling the tight interaction between the gas motions and the
nuclear activity (see e.g. the case of NGC~4388, where we have shown
that the gas is certainly connected to the intracluster gas and that
the source of ionization is not coincident with the optical nucleus, 
Petitjean \& Durret 1993).  In such studies, the knowledge of the host
galaxy morphology is crucial since it strongly constrains the
framework in which the observations should be analyzed (presence of a
companion, distorsion of the disk, orientation and inclination of the
galaxy).

A number of extended ionized nebulae have been detected with radii of
the order of tens to hundreds of kpc from the nucleus itself (see
catalogues by Durret 1989, and Heckman et al. 1991). The properties of
these gaseous nebulae appear to be tightly correlated with the AGN
activity, in particular with the UV and radio emission (McCarthy et
al. 1987, Durret 1990), and the gas often suffers from important
turbulent motions (Jors\"ater et al. 1984, Bergeron et al. 1989,
Heckman et al. 1991, Durret et al. 1994). Besides, the emission line
widths are observed to increase with redshift (Heckman et al. 1991),
implying that the properties of the gaseous nebulae evolve with
redshift. For $z\approx$2 the properties of the ionized gas around
quasars and radio galaxies seem to be similar, as would be the case
for similar objects seen from different viewing angles (Barthel 1989).
However, an analysis of the IRAS data for a sample of radio quasars
and radio galaxies has shown a systematic difference between the IR
properties of these two types of objects (Heckman et al. 1992).

The morphologies of the host galaxies have been investigated by
Hutchings (1987) through optical broad band images of radio galaxies
and radio quasars; he concludes that 80\% of them are interacting.
However, for objects with redshifts of about 0.3, the
[OIII]$\lambda$500.7 emission line is redshifted in the R band and can
contribute most of the light. A striking example is TON~616 for which
the [OIII]$\lambda$500.7 map (Durret et al. 1994) perfectly matches
the R band image (Hutchings \& McClure 1990). Near infrared imaging is
not sensitive to young stellar populations, but on the contrary allows
to sample mainly the old stellar population and can reveal
unambiguously the structure of the underlying host. Moreover, near
infrared images with a good spatial resolution in the K band are
particularly well adapted to this purpose, since at these wavelengths
the contributions of both dust and young ionizing stars are also
minimized. The K band images will reveal the morphologies and near
environments of the harboring galaxies, and allow us to connect the
emitting gas morphology and kinematics (when available) with the
potential of the underlying galaxy. In particular, such data should
allow to study what kinds of perturbations could be invoked as the
origin of the gas fuelling (presence of close companions,
nonaxisymmetrical perturbations in accretion or merging events, or
tidal interacting forces).

We present here observations in the J and K' bands of a sample of
eleven quasars that were known to be associated with an extended
ionized nebulosity. Three other nearby quasars (indicated with an $^*$
in Table 1) were included because they had extensions in their optical
images which could be contaminated by ionized gas emission
lines. Finally, A~0401-350A was observed serendipitously because it
was in our redshift range and there was no object in our sample at
this right ascension.

\section{The data }\label{data}

\begin{table*}[h!]
\caption[ ]{Journal of Observations. }
\begin{tabular} {lrrccrlllc}
\\
\hline
\\
Object       &Coordinates& Observing & Filter & Scale$^a$ & Exposure & Seeing 
& Source     & $\mu_{2\sigma}$ & Radio \\
name         &           &    date~~ &        &       & time~~~ & FWHM 
& redshift   & (mag/ & \\
             &           &           &        &       & (minutes) & (arcsec) 
&            & arcsec$^2$) & \\
\\
\hline
\\
A~0401-350A     & $0401-35$ &  1/2/96 & J & C & 30 & 1.20 & 0.22 &22.0  & No\\
PKS~0812+020    & $0812+02$ & 31/1/96 & J & C & 30 & 1.34 & 0.402 &22.1 & Yes\\
PKS~0812+020    & $0812+02$ & 31/1/96 & J & B & 45 & 1.03 &  &21.0 & \\
PKS~0812+020    & $0812+02$ & 31/1/96 & K' & B & 45 & 1.00 &  &19.7 & \\
PKS~0837-120$^*$    & $0837-12$ &  8/5/96 & J & C & 60 & 1.49 & 0.1976 &22.6 & Yes\\
PKS~0837-120$^*$    & $0837-12$ &  9/5/96 & K' & B & 48 & 1.03 &  &20.3 & \\
3C~215          & $0903+16$ &  1/2/96 & J & C & 30 & 1.30 & 0.411 &22.0 & Yes\\
IRAS~09149-6206$^*$ & $0914-62$ &  1/2/96 & J & C & 3  & 0.96 & 0.057 &20.8 & No \\
IRAS~09149-6206$^*$ & $0914-62$ &  1/2/96 & K' & B & 7  & 0.81 &  &18.8 & \\
PKS~1011-282    & $1011-28$ & 31/1/96 & J & C & 45 & 1.10 & 0.253 &21.9 & Yes \\
3C~275.1        & $1241+16$ &  8/5/96 & J & C & 36 & 1.62 & 0.557 &22.6 & Yes\\
3C~275.1        & $1241+16$ &  9/5/96 & K' & B & 48 & 1.31 &  &20.3 & \\
PKS~1302-102$^*$    & $1302-10$ & 31/1/96 & J & C & 30 & 1.20 & 0.286 &21.8 & Yes \\
3C~281          & $1305+06$ &  8/5/96 & J & C & 30 & 1.32 & 0.599 &22.7 & Yes \\
3C~281          & $1305+06$ &  8/5/96 & K' & B & 42 & 1.36 &  &20.2 & \\
4C~20.33        & $1422+20$ &  9/5/96 & J & C & 36 & 1.52 & 0.871 &22.6 & Yes \\
4C~11.50        & $1548+11$ &  9/5/96 & J & C & 30 & 1.52 & 0.436 &22.5 & Yes \\
4C~11.50        & $1548+11$ &  9/5/96 & K' & B & 42 & 1.11 &  &20.1 & \\
MRK 877         & $1617+17$ &  8/5/96 & J & C & 30 & 1.42 & 0.114 &22.7 & No\\
3C~334          & $1618+17$ &  8/5/96 & J & C & 30 & 1.37 & 0.555 &22.6 & Yes \\
MC~1745+163     & $1745+16$ &  8/5/96 & J & C & 30 & 1.29 & 0.392 &22.4 & Yes \\
MC~1745+163     & $1745+16$ &  9/5/96 & K' & B & 48 & 1.31 &  &20.3 & \\
4C~11.72        & $2251+11$ &  8/5/96 & J & C & 42 & 1.47 & 0.323 &22.4 & Yes\\
4C~11.72        & $2251+11$ &  9/5/96 & K' & B & 36 & $^{b}$     &  &19.9 & \\
\\
\hline
\end{tabular}
\begin{footnotesize}

$^a$ C=0.507''/pix, B=0.278''/pix\\
$^b$ No star in the frame
\end{footnotesize}

\protect\label{datatab}
\end{table*}

The data were obtained during two runs with the ESO 2.2m telescope
with IRAC2B in 1996. The detector was a NICMOS-3 array with pixels of
size 40$\mu$m, a Read Out Noise of 46 e$^-$ and a Gain of 6.6
e$^-$/ADU. We used lenses B (0.278''/pix) and C (0.507''/pix), that
give field sizes of 71''x71'' and 129''x129'' respectively.  The
observing strategy for each object was the following: we first
obtained a J image with lense C and if it showed any hint of an
extension we changed to lense B and obtained the K' image.  Details on
the observations are given in Table \ref{datatab}.  For every target,
we obtained a number of images (less than 1 minute exposure each)
shifting the quasar position by about 15''-20'' until the total
exposure time was attained.  For the flux calibration we measured the
following standard stars: HD29250, HD56189, HD84503, HD105116 (first
run), HD84090, HD1148951, DM597287, HD177619 and SJ9149 (second
run). The data reduction and calibration was performed, following the
standard procedures with the IRAF\footnote{IRAF is the Image Analysis
and Reduction Facitily made available to the astronomical community by
the National Optical Astronomy Observatories, which are operated by
the Association of Universities for Research in Astronomy (AURA),
Inc., under contract with the U.S. National Science Foundation.}
software and SQIID package. We used dome flat-fields by subtracting
the images taken with the lamps off from those taken with the lamps
on. We compute the median of object images to obtain the corresponding
sky frame. Sky-subtracted images were used to obtain the final
mosaics.  We reach a photometric accuracy of about 10\%. In column 9
of Table 1 we give the isophotal magnitude/arcsec$^2$ corresponding to
2$\sigma$ of the background for the regions close to the quasar (where
the S/N is the highest). Total magnitudes given in Table 2 were
measured in circular apertures.

Deconvolution from the point spread function was performed using the
Lucy algorithm in the stsdas.analysis.restore package. We estimated
the shape of the PSF by using starlike objects in the field. Average
isophotal profiles were determined using the IRAF task {\bf ellipse}
in the stsdas.analysis.isophote\footnote{STSDAS is distributed by the
Space Telescope Institute, which is operated by AURA, Inc., under NASA
contract NAS~5-26555.} package. 
%When we obtain a suitable PSF after
%deconvolution we calculate the magnitude of the host galaxy and give
%it in Section 3 together with the individual object description.
We have computed the host contribution by subtracting the PSF profile
(for the images with suitable stars to compute the PSF) from that of
the quasar, forcing that the resulting profile has no central hole
(Aretxaga et al. 1995; R\"onnback et al. 1996); the magnitudes of the
host galaxies are given in Table \ref{tabhost}.  Since the
deconvolution procedure does not conserve calibrated fluxes, we have
deconvolved sky-subtracted non-calibrated images. This explains why
the deconvolved images are given with fluxes in arbitrary units
(Figs. 2, 12, 14 and 15).

Table \ref{autour} lists for every object detected in our images: (1)
our order number (mentioned in the text with a \# symbol); (2) and (3)
abridged $\alpha$ and $\delta$ coordinates; (4) and (5) $\alpha$ and
$\delta$ positions relative to the quasar (the quasar coordinates were
taken from the Simbad database, in order to have more accurate values
than those written in the image headers); (6) and (7) our J and K'
magnitude measurements (note that the spatial coverage is smaller for
the higher resolution scale of the K' images); (8) R magnitudes from
published optical photometry; (9) references; (10) numbering in the
corresponding reference; (11) type of the object (1=galaxy, 2=probable
galaxy, 3=star) given by the reference or, when no reference was
given, type assigned by us.  When available, we also give the redshift
in column 11 in parentheses, with the corresponding reference; (12)
Yes for radio-loud, No for radio-quiet objects.

A few of the objects closest to the quasar are indicated in the
various plots (Figures 1 to 16) with their corresponding numbers as
given in Table \ref{autour}, column 1. Whenever they are mentioned in
the text we use the \#{\it number}; when we mention the numbering
given by other authors we just use the {\it number}.

We compare the infra-red images with optical HST images (Figures 17 to
23) from the public archive when available to verifiy in particular
that the extensions that we detect are not a consequence of the
presence of close companions.

\begin{figure}
\centerline{\psfig{figure=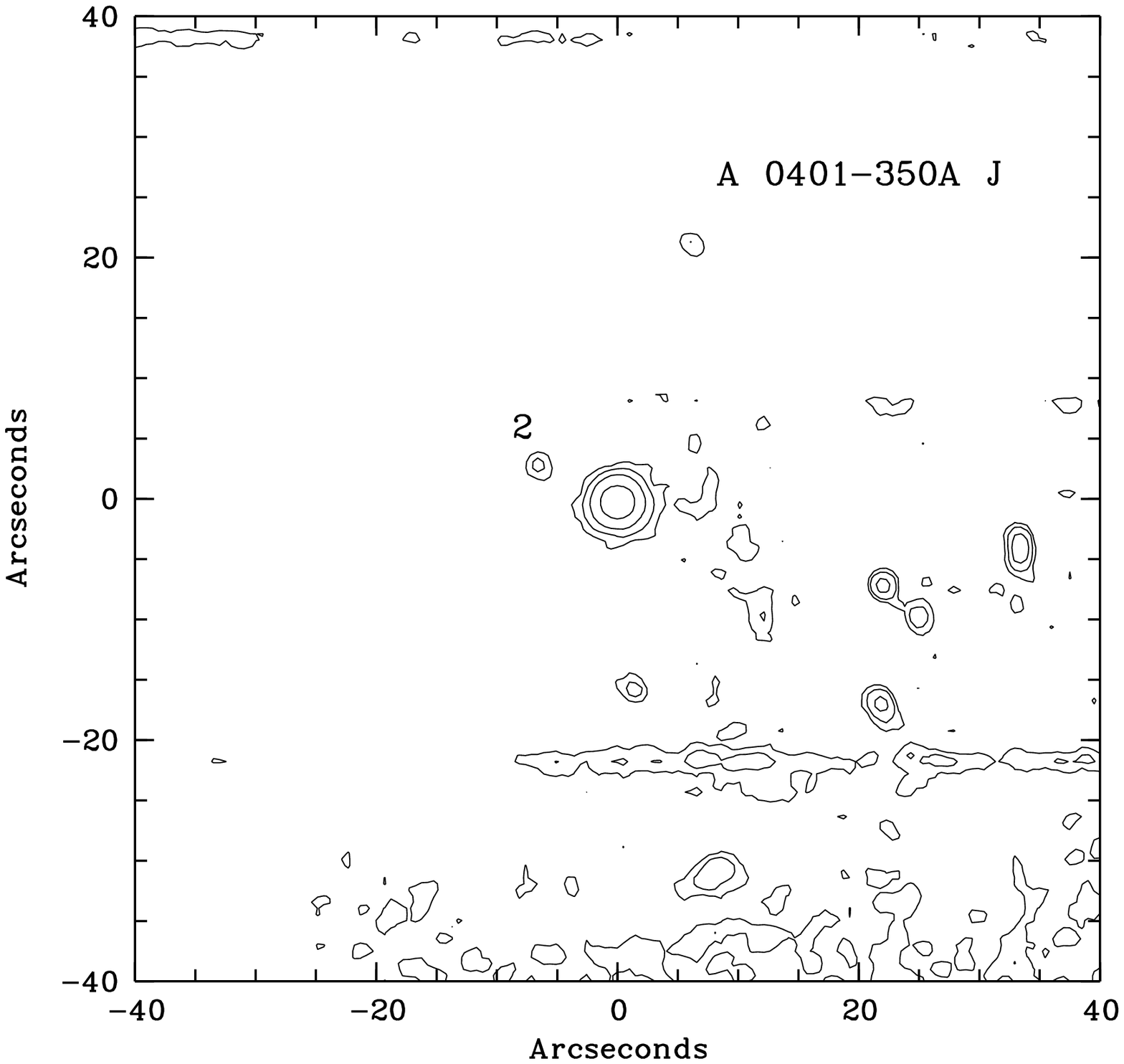,height=6cm}}
\caption[ ]{Image of A~0401-350A in the J band. As in all following figures,
the quasar is at the (0,0) position; North is to the top and East to the left.
Numbers on the figures refer to the object numbers in Table 2. Contour levels
are 2.2 10$^{-10}$, 1.8 10$^{-9}$, 5.7 10$^{-9}$ and 2.9 10$^{-8}$
%6 10$^{-10}$, 1 10$^{-9}$, 2 10$^{-9}$ and 8 10$^{-9}$ 
erg s$^{-1}$ arcsec$^{-2}$.}
\protect\label{a04j}
%\end{figure}
%\begin{figure}

\vspace {0.5truecm}

\centerline{\psfig{figure=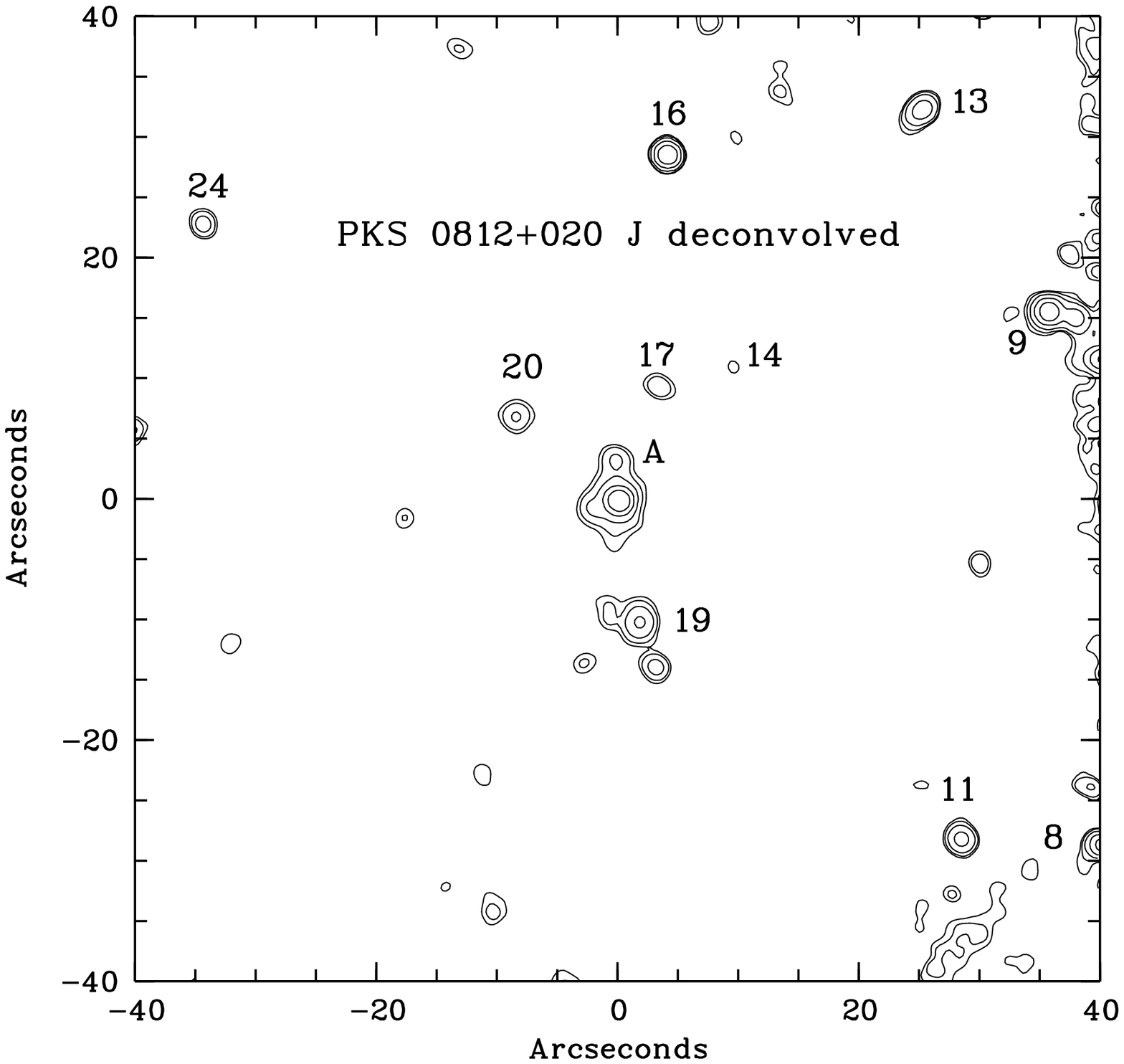,height=6cm}}
\caption[ ]{Image of PKS~0812+020 in the J band after deconvolution for 
seeing effects. Contour levels are 0.075, 0.1, 0.2, 1 and 10 (in arbitrary
units). }
\protect\label{p0812j}
%\end{figure}
%\begin{figure}

\vspace {0.5truecm}

\centerline{\psfig{figure=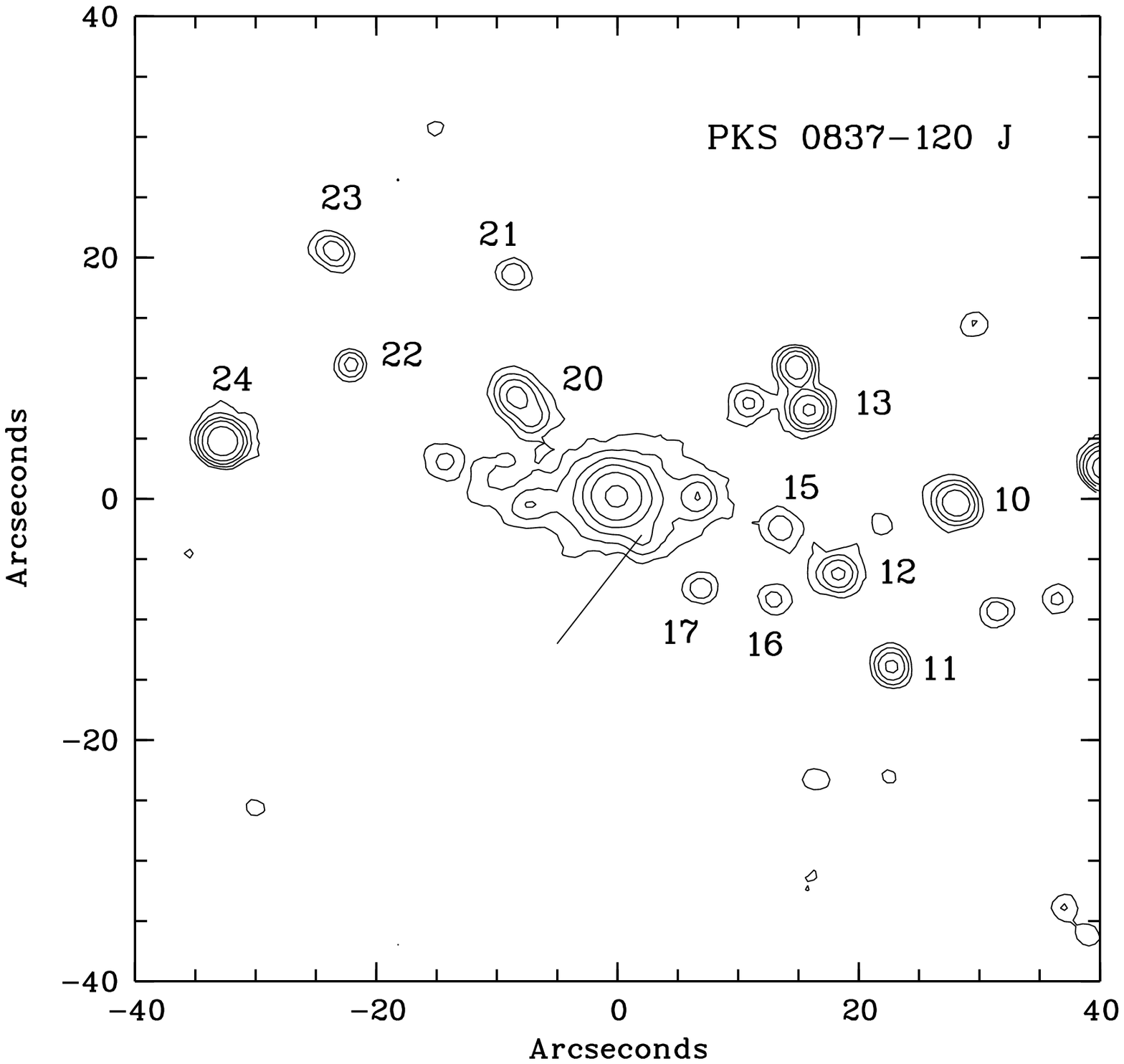,height=6cm}}
\caption[ ]{Image of PKS 0837-120 in the J band. Contour levels are
1.0 10$^{-9}$, 2.2 10$^{-9}$, 4.1 10$^{-9}$, 8.0 10$^{-9}$, 2.0 10$^{-8}$
and 1.2 10$^{-7}$ 
%2 10$^{-10}$, 5 10$^{-10}$, 1 10$^{-9}$, 2 10$^{-9}$, 5 10$^{-9}$, 3 10$^{-8}$
erg s$^{-1}$ arcsec$^{-2}$.}
\protect\label{p0837j}
\end{figure}
\begin{figure}

\centerline{\psfig{figure=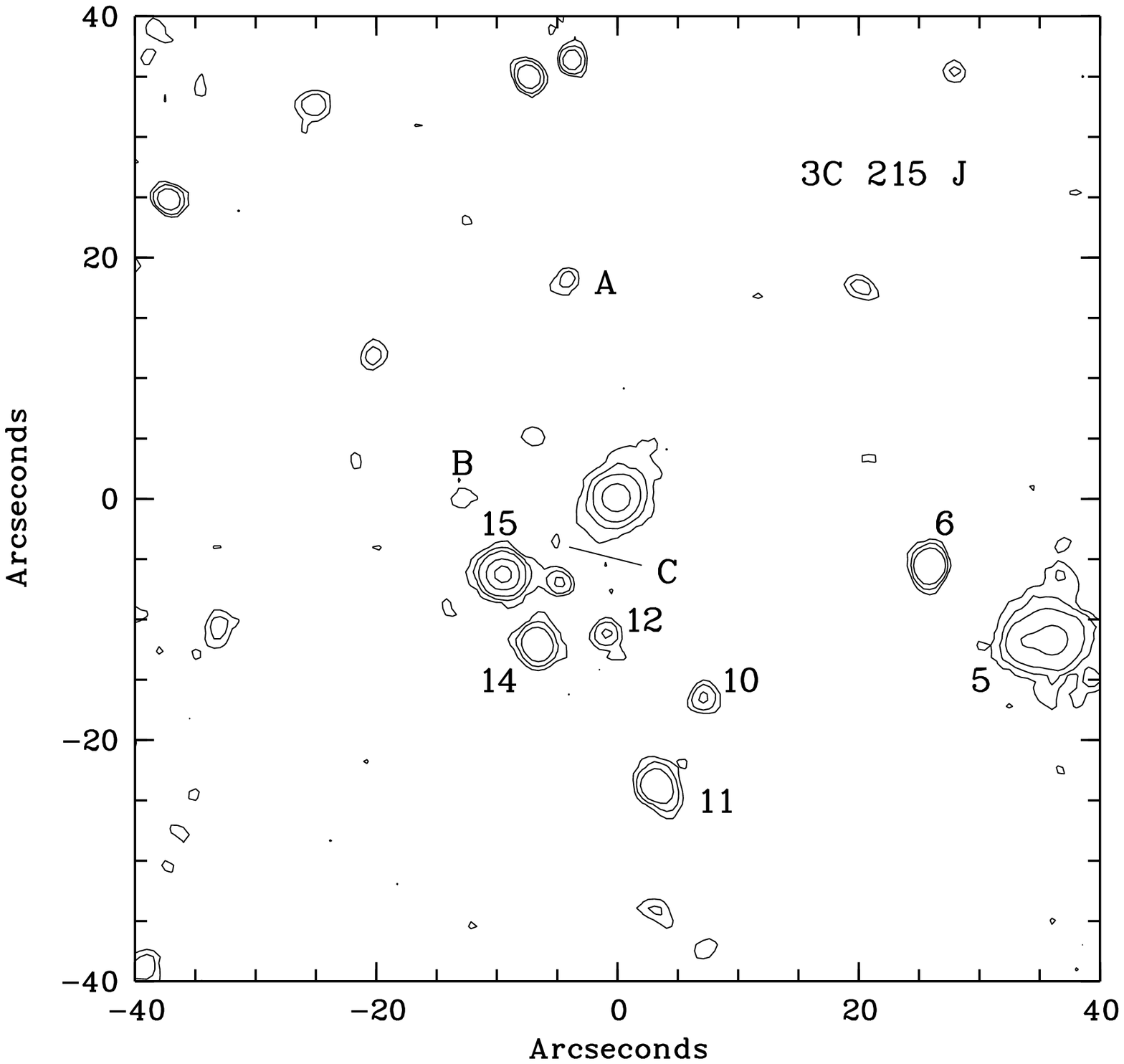,height=6cm}}
\caption[ ]{Image of 3C~215 in the J band. Contour levels are
2.4 10$^{-9}$, 3.4 10$^{-9}$, 5.3 10$^{-9}$, 2.1 10$^{-8}$ and 7.9 10$^{-8}$
%2.5 10$^{-10}$, 5 10$^{-10}$, 1 10$^{-9}$, 5 10$^{-9}$, and 2 10$^{-8}$ 
erg s$^{-1}$ arcsec$^{-2}$.}
\protect\label{3c215j}
%\end{figure}
%\begin{figure}

\vspace {0.5truecm}

\centerline{\psfig{figure=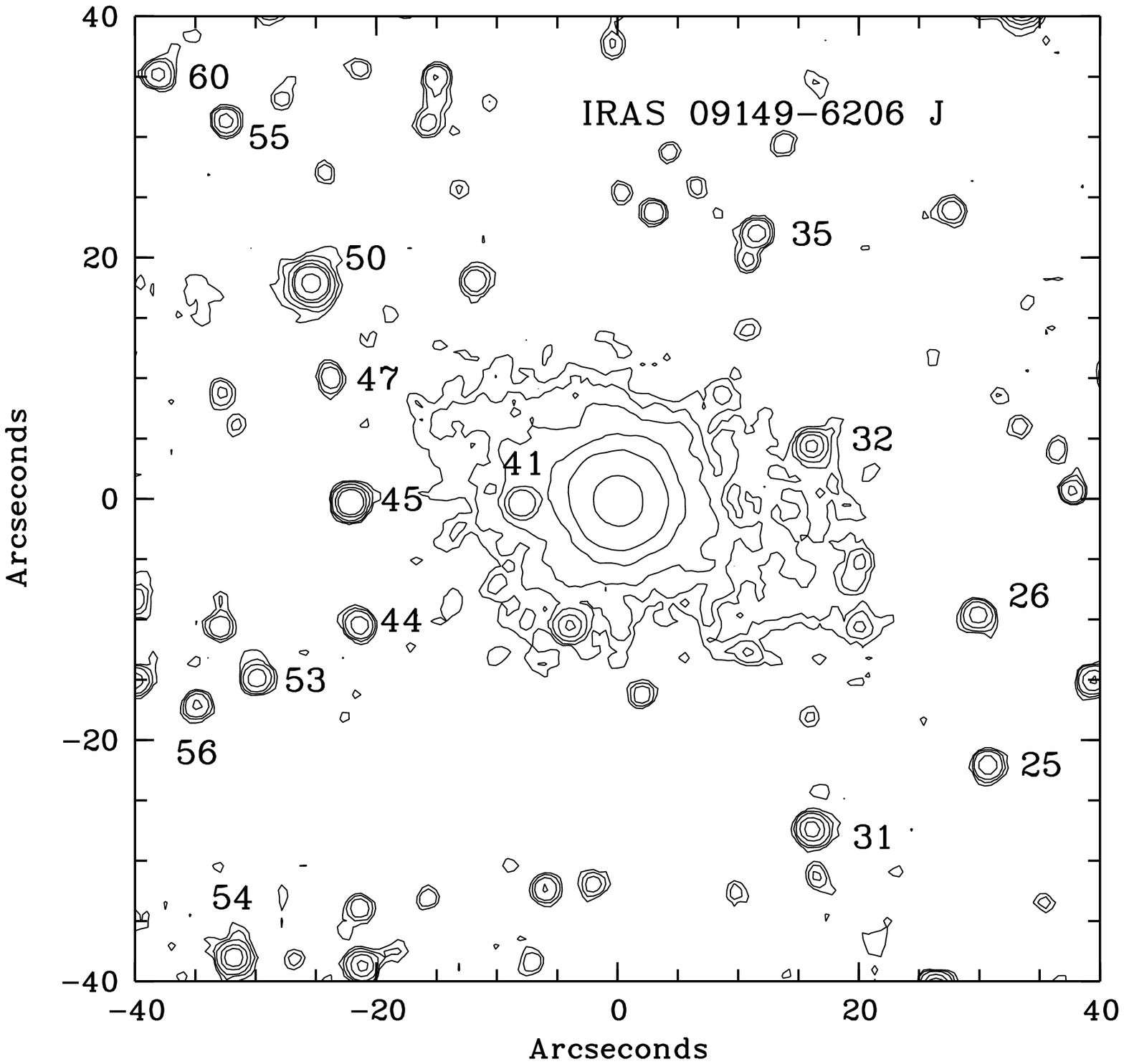,height=6cm}}
\caption[ ]{Image of IRAS~09149-6206 in the J band. Contour levels are
1.7 10$^{-8}$, 1.9 10$^{-8}$, 2.3 10$^{-8}$, 3.5 10$^{-8}$, 5.4 10$^{-8}$ and
2.1 10$^{-7}$
%5 10$^{-10}$, 1 10$^{-9}$, 2 10$^{-9}$, 5 10$^{-9}$, 1 10$^{-8}$ and 
%5 10$^{-8}$ 
erg s$^{-1}$ arcsec$^{-2}$.}
\protect\label{i09j}
%\end{figure}
%\begin{figure}

\vspace {0.5truecm}

\centerline{\psfig{figure=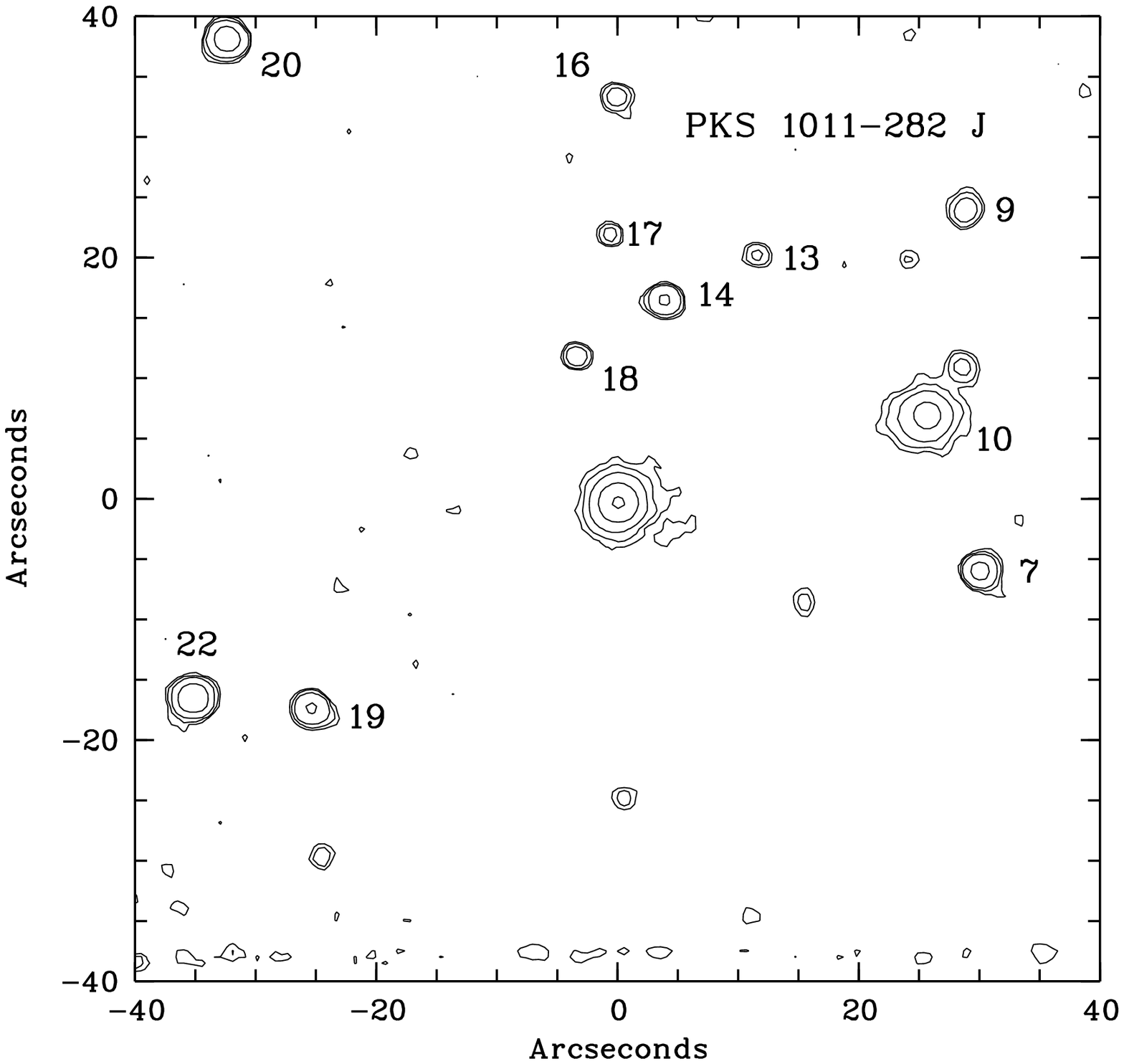,height=6cm}}
\caption[ ]{Image of PKS~1011-282 in the J band. Contour levels are
9.6 10$^{-10}$, 1.7 10$^{-9}$, 3.7 10$^{-9}$, 1.9 10$^{-8}$ and 1.9 10$^{-7}$ 
%3 10$^{-10}$, 5 10$^{-10}$, 1 10$^{-9}$, 5  10$^{-9}$ and 5  10$^{-8}$
erg s$^{-1}$ arcsec$^{-2}$.}
\protect\label{p1011j}
\end{figure}
\begin{figure}
\centerline{\psfig{figure=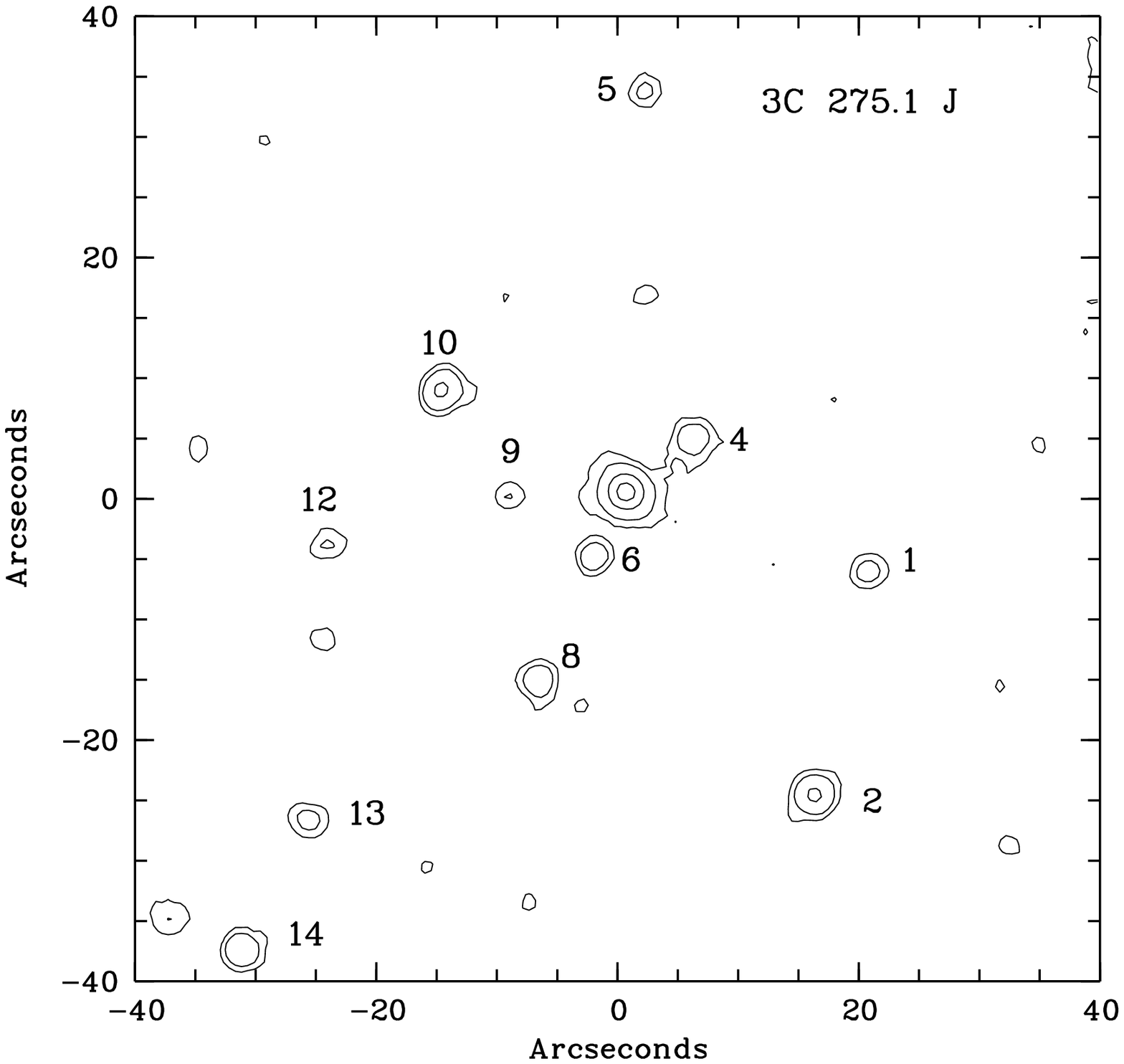,height=6cm}}
\caption[ ]{Image of 3C~275.1 in the J band. Contour levels are
1.0 10$^{-9}$, 1.7 10$^{-9}$, 8.0 10$^{-9}$ and 2.0 10$^{-8}$
%2 10$^{-10}$, 5 $^{-10}$, 2 $^{-9}$ and 5 $^{-9}$
erg s$^{-1}$ arcsec$^{-2}$.}
\protect\label{3c275j}
%\end{figure}
%\begin{figure}

\vspace {0.5truecm}

\centerline{\psfig{figure=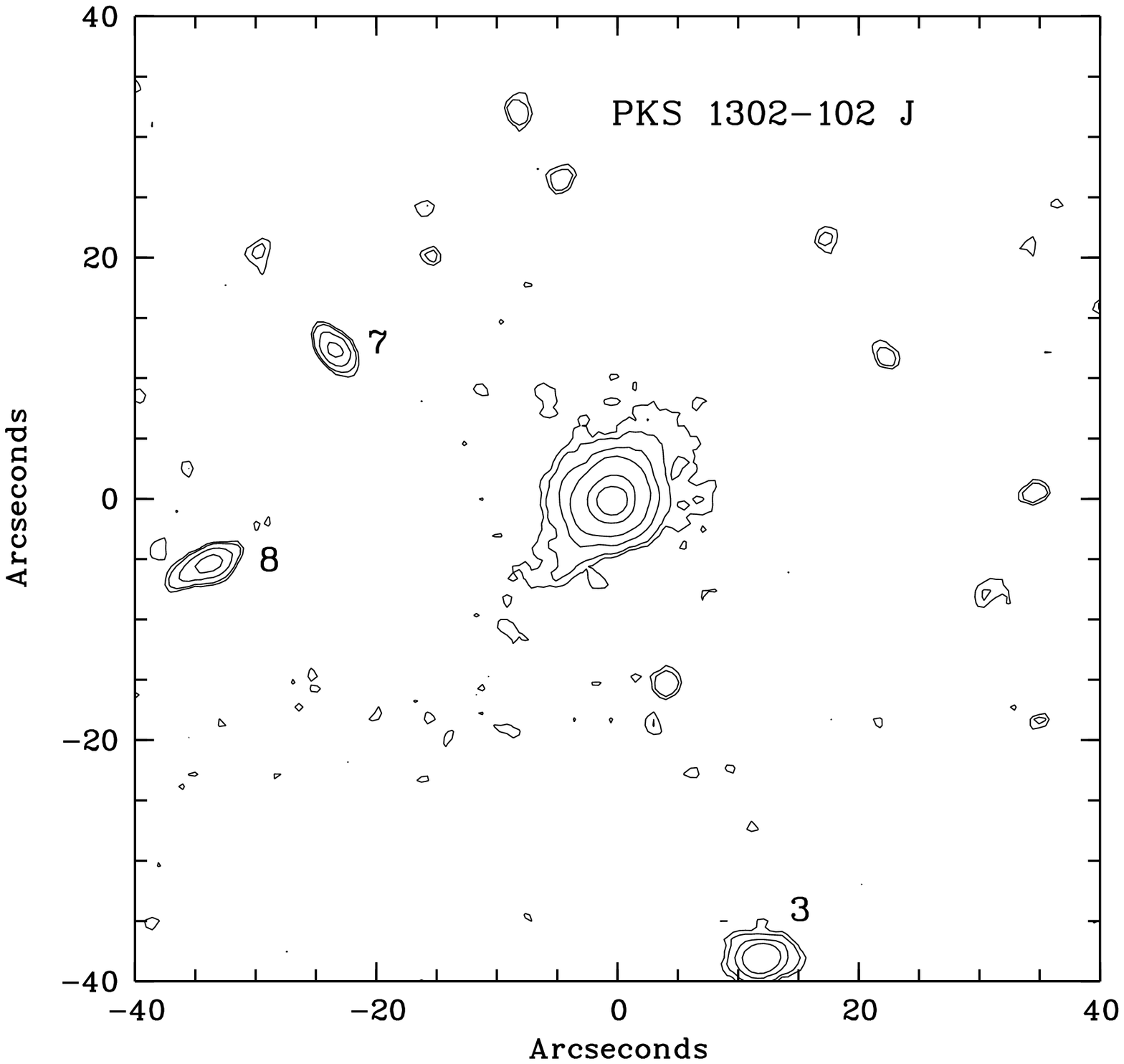,height=6cm}}
\caption[ ]{Image of PKS~1302-102 in the J band. Contour levels are
3.0 10$^{-10}$, 8.9 10$^{-10}$, 2.8 10$^{-9}$, 6.7 10$^{-9}$, 3.8 10$^{-8}$
and 1.9 10$^{-7}$
%3.5 10$^{-10}$, 5 10$^{-10}$, 1 10$^{-9}$, 2 10$^{-9}$, 1 10$^{-8}$ and
%5 10$^{-8}$ 
erg s$^{-1}$ arcsec$^{-2}$.}
\protect\label{p1302j}
%\end{figure}
%\begin{figure}

\vspace {0.5truecm}

\centerline{\psfig{figure=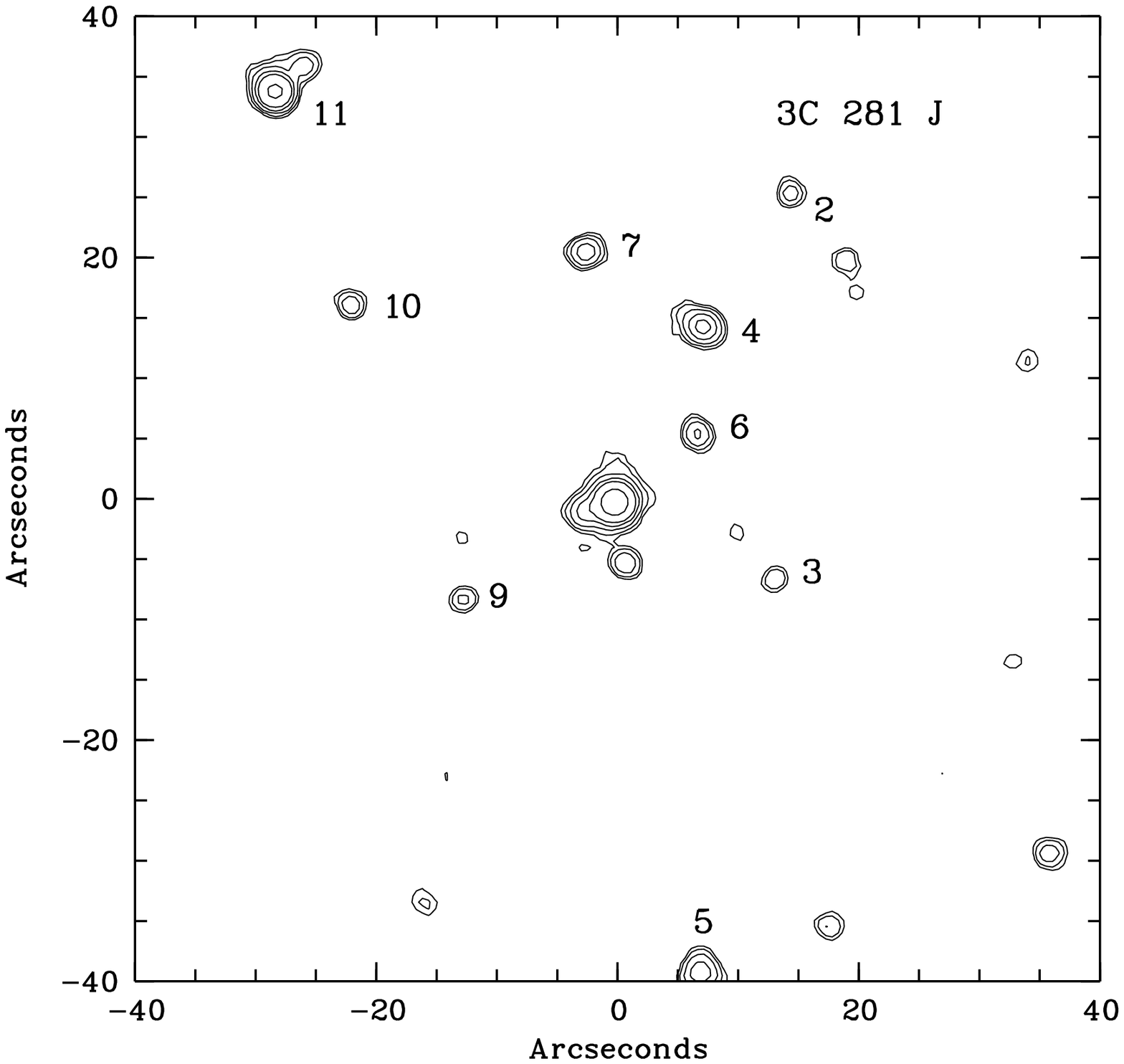,height=6cm}}
\caption[ ]{Image of 3C~281 in the J band. Contour levels are
9.9 10$^{-10}$, 1.4 10$^{-9}$, 2.2 10$^{-9}$, 3.7 10$^{-9}$, 2.9 10$^{-9}$
and 2.2 10$^{-8}$
%$-4\ 10^{-10}$, $-3\ 10^{-10}$, $-1\  10^{-10}$, 3 $10^{-10}$, 1 $10^{-10}$
%and 5 $10^{-9}$ 
erg s$^{-1}$ arcsec$^{-2}$. }
\protect\label{3c281j}
\end{figure}
\begin{figure}
\centerline{\psfig{figure=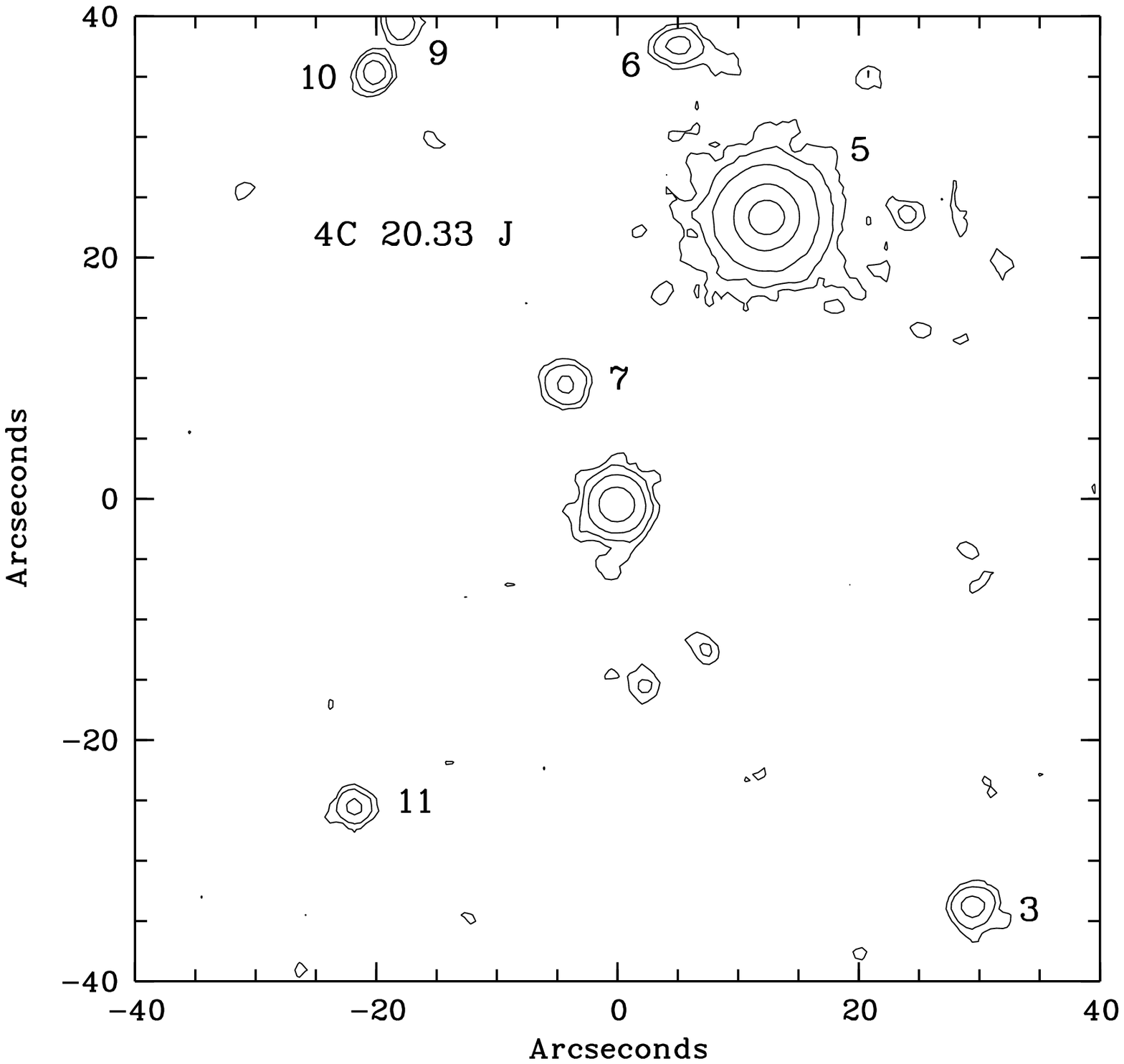,height=6cm}}
\caption[ ]{Image of 4C~20.33 in the J band. Contour levels are
6.6 10$^{-10}$, 1.2 10$^{-9}$, 3.2 10$^{-9}$, 1.9 10$^{-8}$ and 1.9 10$^{-7}$
%3.5 10$^{-10}$, 5 10$^{-10}$, 1 10$^{-9}$, 5 10$^{-9}$ and 5 10$^{-8}$
erg s$^{-1}$ arcsec$^{-2}$.}
\protect\label{4c20j}
%\end{figure}
%\begin{figure}

\vspace {0.5truecm}

\centerline{\psfig{figure=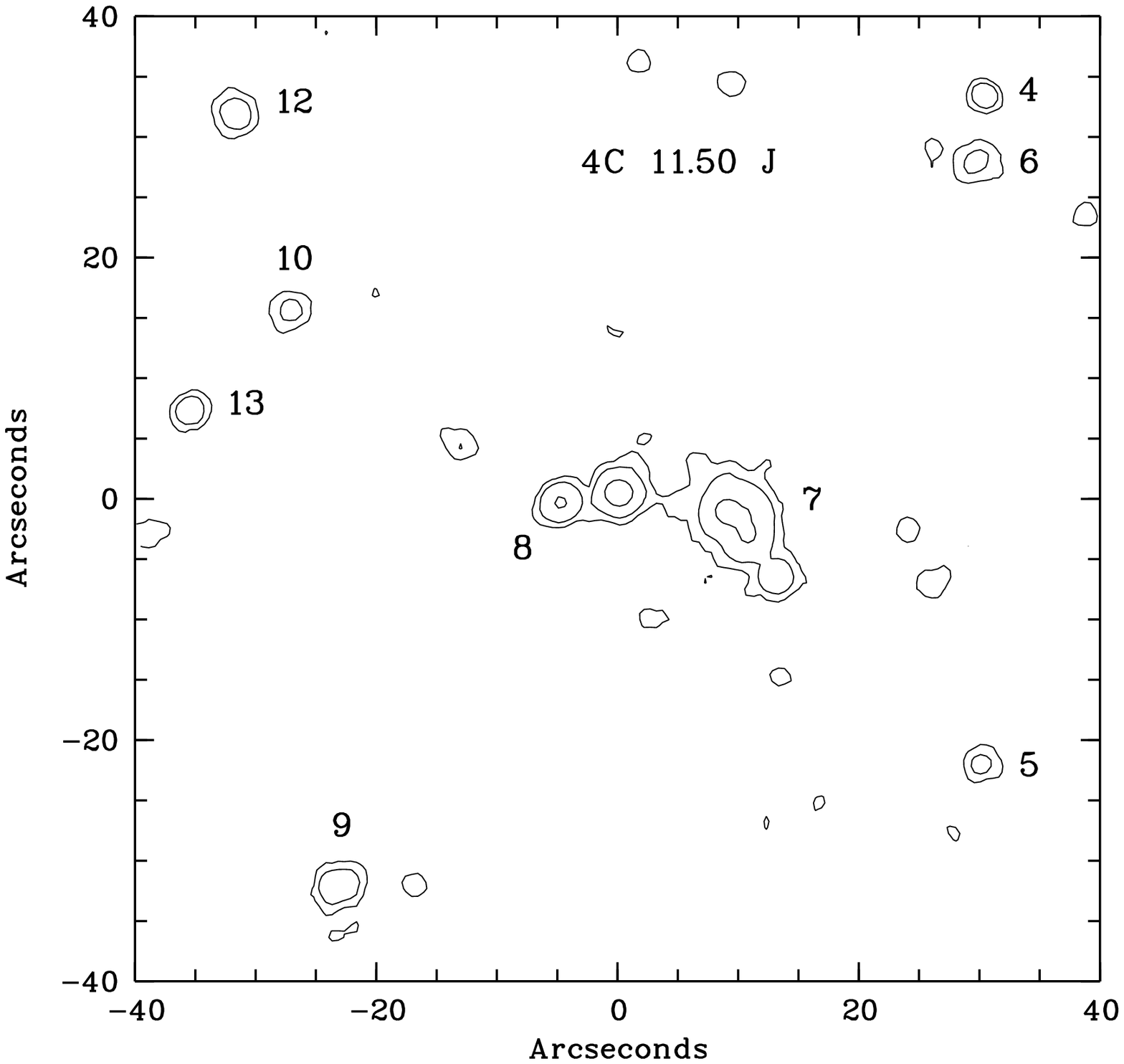,height=6cm}}
\caption[ ]{Image of 4C 11.50 in the J band. Contour levels are
8.9 10$^{-10}$, 2.1 10$^{-9}$ and 9.8 10$^{-9}$
%7 10$^{-10}$, 1 10$^{-9}$ and 3 10$^{-9}$ 
erg s$^{-1}$ arcsec$^{-2}$.}
\protect\label{4c1150j}
%\end{figure}
%\begin{figure}

\vspace {0.5truecm}

\centerline{\psfig{figure=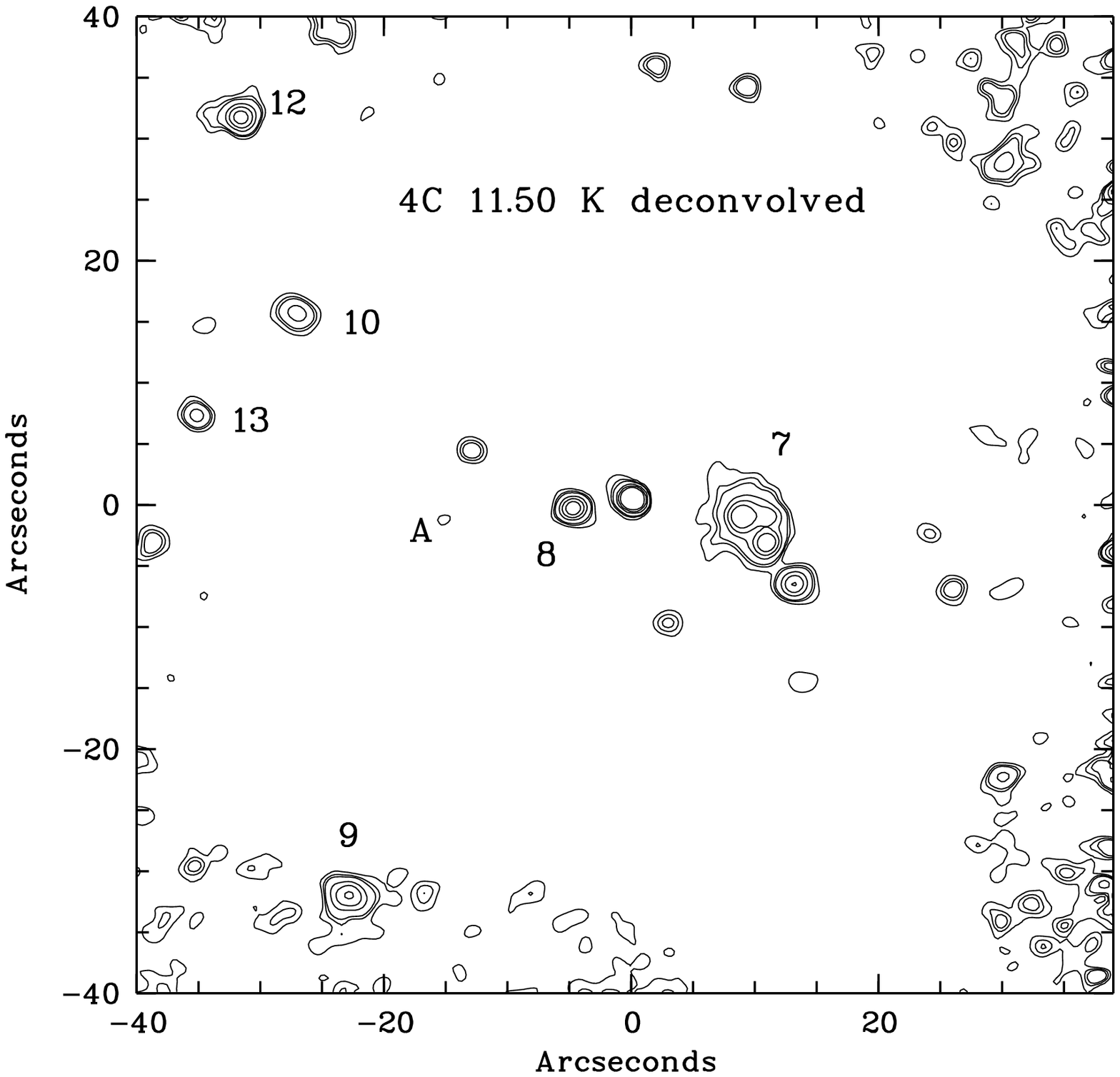,height=6cm}}
\caption[ ]{Image of 4C 11.50 in the K' band corrected for seeing effects.
Contour levels are 1, 1.5, 2, 5, 10 and 20 (in arbitrary units).}
\protect\label{4c1150k}
\end{figure}
\begin{figure}
\centerline{\psfig{figure=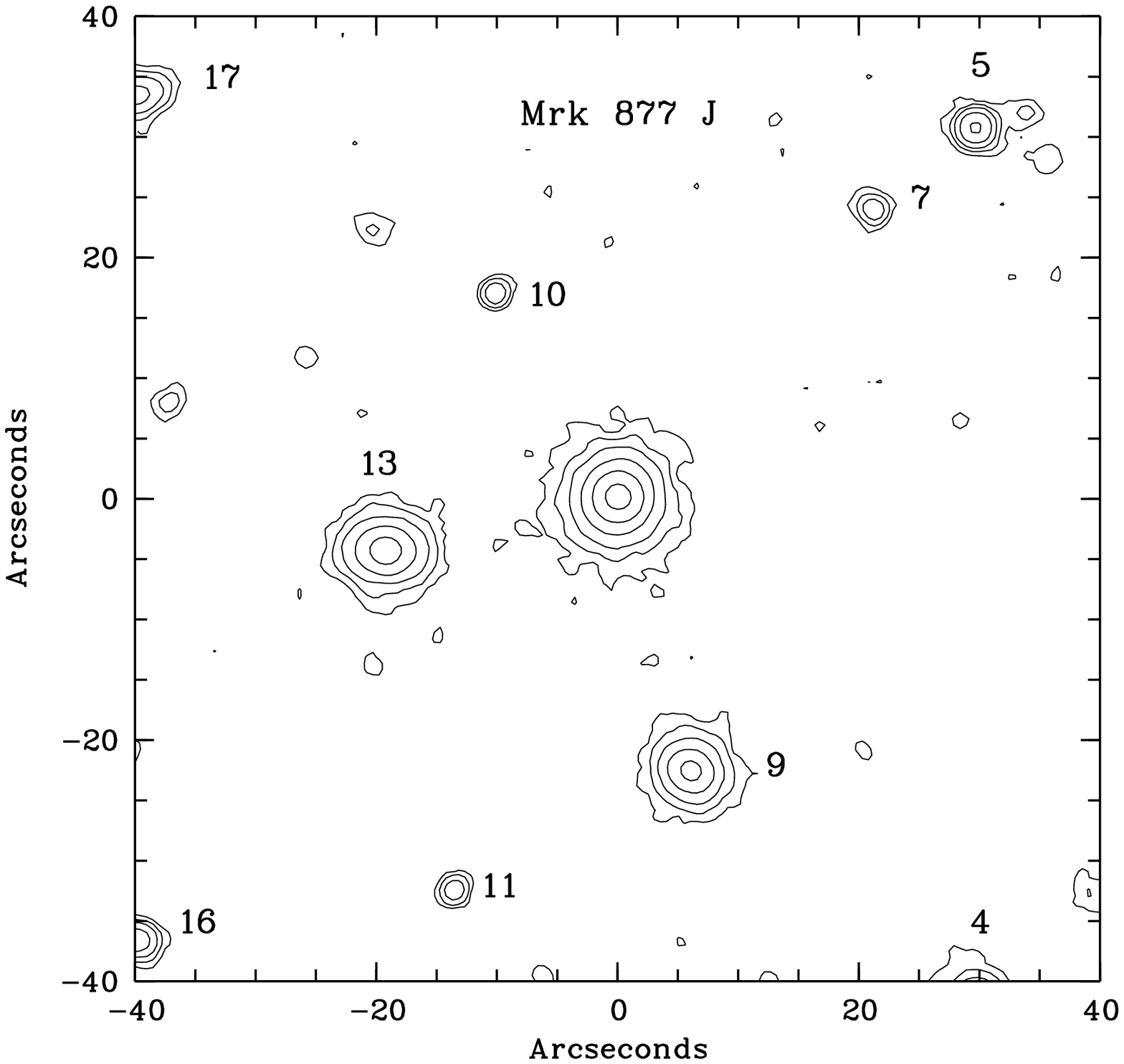,height=6cm}}
\caption[ ]{Image of MRK~877 in the J band. Contour levels are
2.8 10$^{-9}$, 3.6 10$^{-9}$, 5.1 10$^{-9}$, 1.0 10$^{-8}$, 3.0 10$^{-8}$
and 2.0 10$^{-7}$
%1 10$^{-10}$, 3 10$^{-10}$, 7 10$^{-10}$, 2 10$^{-9}$, 7 10$^{-9}$ and
%5 10$^{-8}$ 
erg s$^{-1}$ arcsec$^{-2}$. }
\protect\label{mk877j}
%\end{figure}
%\begin{figure}

\vspace {0.5truecm}

\centerline{\psfig{figure=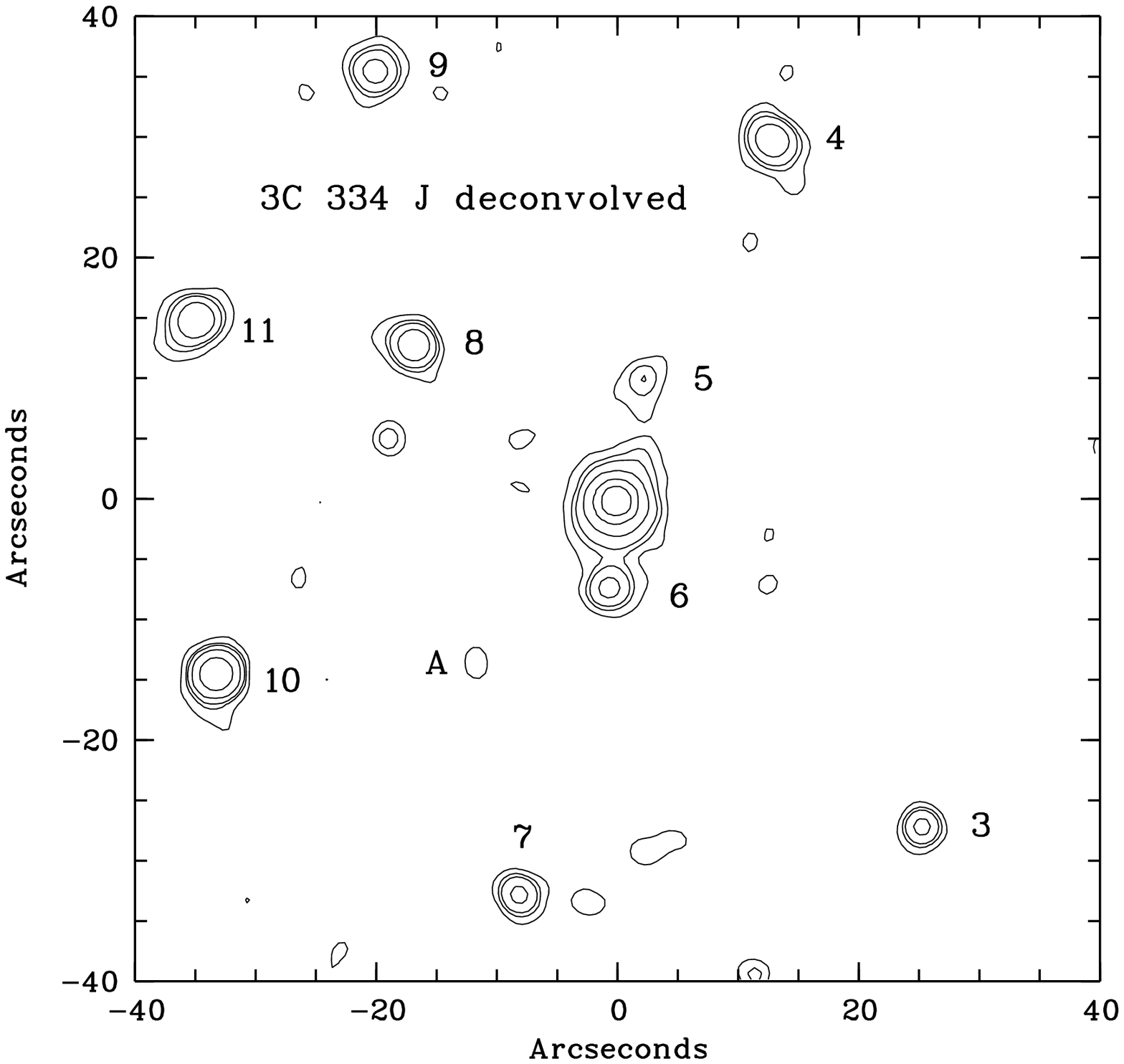,height=6cm}}
\caption[ ]{Image of 3C~334 in the J band after deconvolution for seeing 
effects. Contour levels are
0.8, 1.0, 1.2, 2.0, 10 and 100 (in arbitrary units).}
\protect\label{3c334j}
%\end{figure}
%\begin{figure}

\vspace {0.5truecm}

\centerline{\psfig{figure=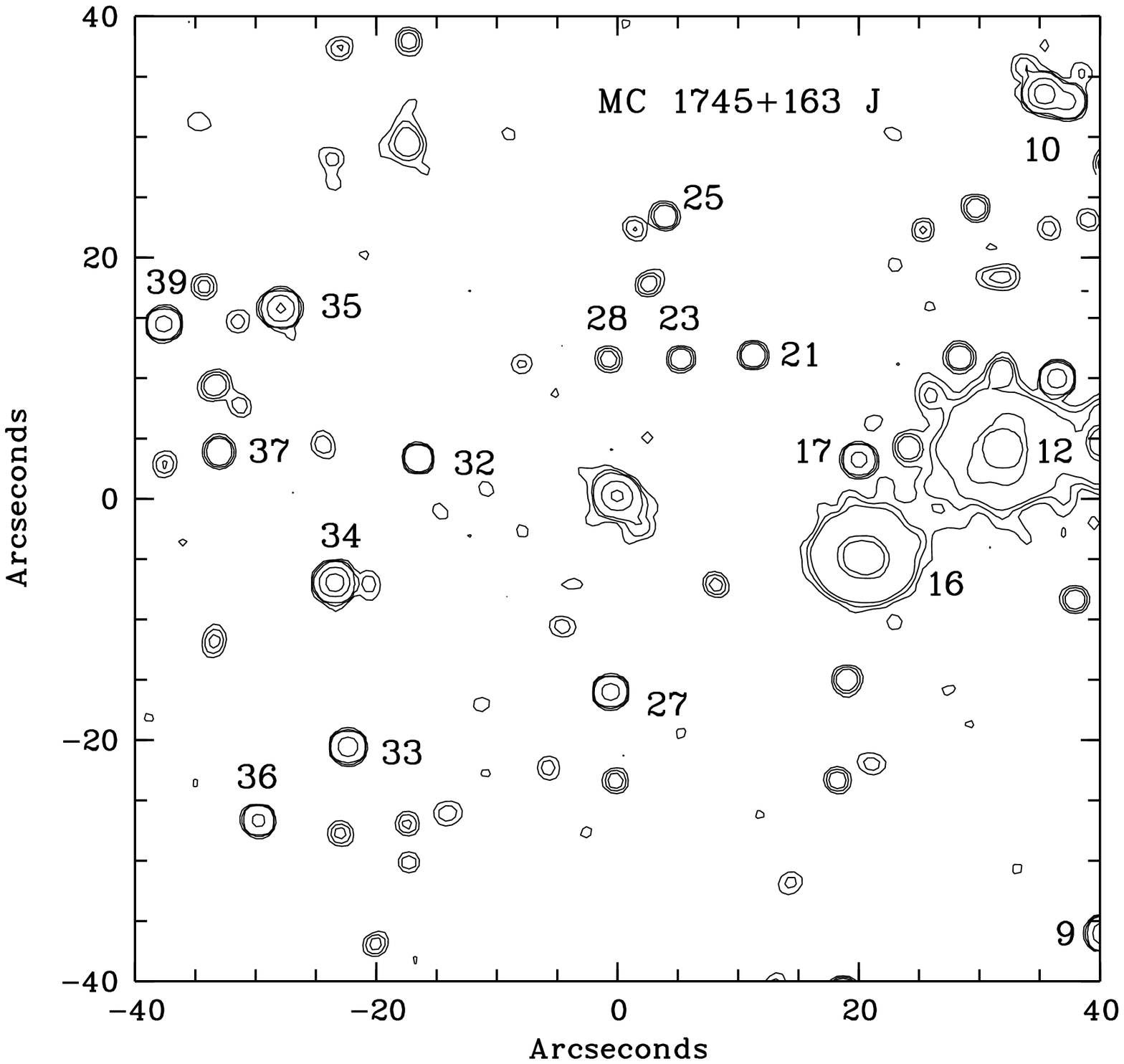,height=6cm}}
\caption[ ]{Image of MC 1745+163 in the J band. Contour levels are
3, 5, 7.5, 50 and 200 (in arbitrary units).}
\protect\label{m1745j}
\end{figure}
\begin{figure}

\vspace {0.5truecm}

\centerline{\psfig{figure=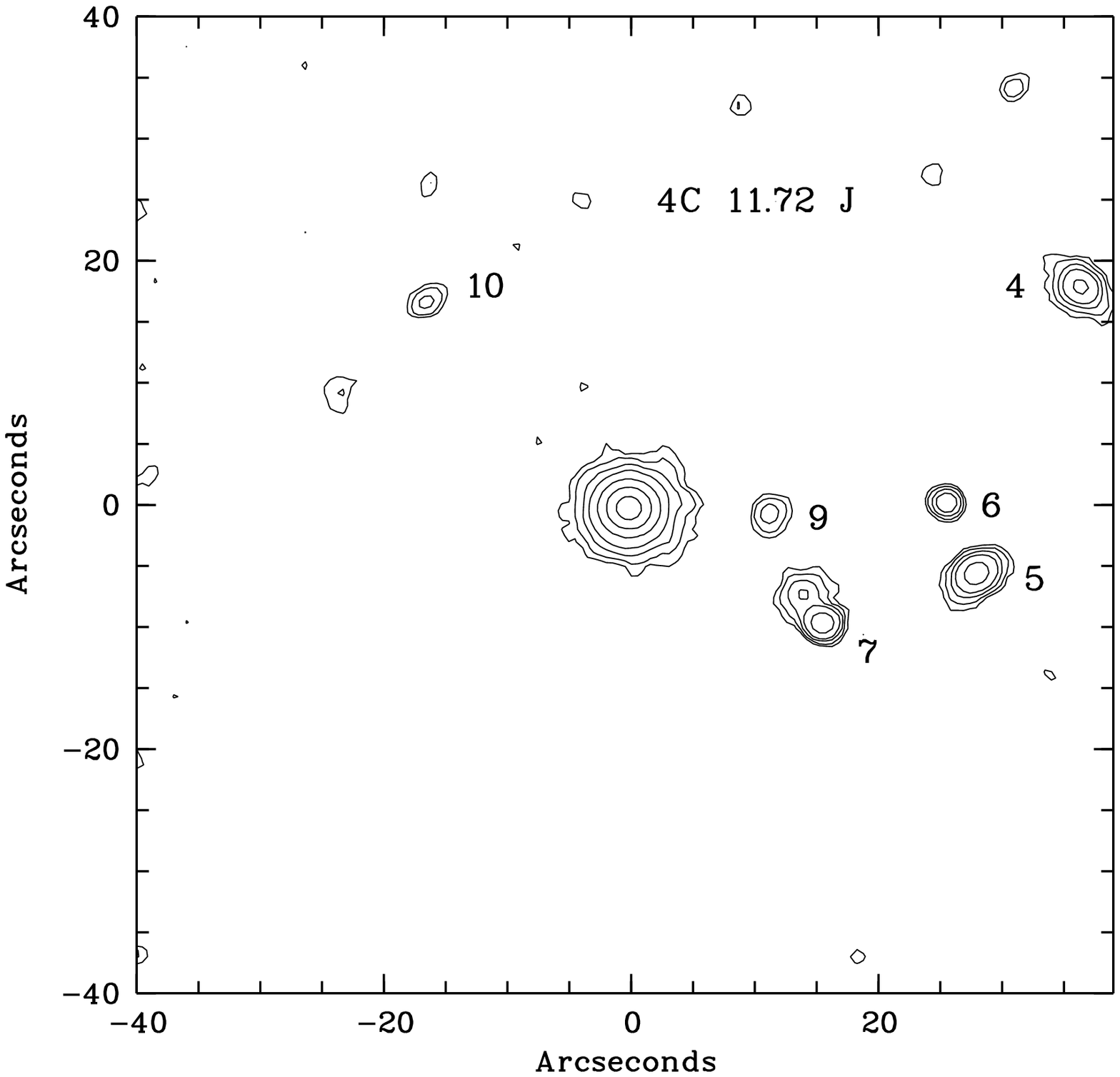,height=6cm}}
\caption[ ]{Image of 4C~11.72 in the J band. Contour levels are
7.4 10$^{-10}$, 1.3 10$^{-9}$, 2.5 10$^{-9}$, 4.4 10$^{-9}$,
1.1 10$^{-8}$, 4.2 10$^{-8}$ and 2.0 10$^{-7}$
%$-6.5\ 10^{-10}$, $-5\ 10^{-10}$, $-2\ 10^{-10}$, 3 $10^{-10}$,
%2 $10^{-9}$, 1 $10^{-8}$ and 5 $10^{-8}$ 
erg s$^{-1}$ arcsec$^{-2}$.}
\protect\label{4c1172j}
\end{figure}

\section{Description of individual objects.}

\subsection{A~0401-350A  (z=0.22)}

Very little is known about this quasar from previous studies.  We do
not detect any extension even after deconvolution and PSF
subtraction. Note however that the background is quite noisy,
especially to the South-East (Fig.~\ref{a04j}); this prevents a
reliable determination of the PSF at faint levels, making the analysis
of the presence of faint extensions very uncertain.

\subsection{PKS~0812+020  (z=0.402)}

According to Hutchings \& Neff (1990), PKS~0812+020 is a one-sided
lobe radio source, with a very close neighbour only 7~kpc away (in
probable tidal interaction) and 16 galaxies within a distance of $\pm
30$".

The J band image shows structures similar to what is seen in the
optical by Wyckoff et al. (1981), in their search for the underlying
galaxy. The radio lobe 10'' to the North-West, found to be coincident
with a diffuse optical emission region (Wyckoff et al. 1983),
corresponds to an object (\#17) of magnitude 18.8 in J and 17.7 in
K'. We detect extensions to the East and South out to about 3'' and
2.5'' respectively from the quasar, and a very close object 3.5'' to
the North (object A). The deconvolved image (Fig. \ref{p0812j}) shows
that the counterpart of the radio lobe (object \#17) has elliptical
contours; however Ellingson et al. (1991a) have not measured its
redshift.

The HST image (Fig. \ref{p0812hst}) shows that the extension seen to
the east in the J band image is an independent object lying 2'' from
the quasar (object B).  The optical counterpart of the radio lobe
(\#17) seems to be a spiral galaxy. The extension to the South is not
visible in this image. From the HST image we can also say that objects
\#14, \#19 and \#20 are galaxies.

In Table \ref{autour}, four objects taken from Ellingson et
al. (1991a) are quoted with redshifts between 0.30 and 0.408.  Out of
these, only objects \#11 and \#19 are inside the field shown in
Fig. \ref{p0812j}. We do not detect galaxy 7 in Ellingson et
al. (1991a), and their object 12 (z=0.4038) is unfortunately outside
our field.  Note that object \#19 is exactly at the same redshift as
the quasar.

\subsection{PKS~0837-120  (3C 206, z=0.1976)}

This quasar is a steep spectrum, classical double radio source (Miley
\& Hartsuijker 1978), strongly variable at optical wavelengths (see
e.g. Ellingson et al. 1989). As suggested by Wyckoff et al. (1981) and
confirmed by Hutchings et al. (1988), the host galaxy of this quasar
can be seen in several wavelength bands. At least three galaxies
appear to be very close to the quasar in projection (V\'eron-Cetty \&
Woltjer 1990), and Ellingson et al. (1989) have confirmed
spectroscopically that PKS~0837-120 is indeed located in a richness
class 1 cluster of galaxies.

Our J band image (Fig. \ref{p0837j}) shows the same structures as
those seen in the optical by V\'eron-Cetty \& Woltjer (1990), with a
number of objects in the close vicinity of the radio quasar. The
deconvolution of the K' image allows to resolve the extension to the
S-SW (indicated with a line on Fig. \ref{p0837j}) as an object; the
host galaxy is also visible and its derived magnitude is
K'$_{gal}$ = 13.7.

In the HST image (Fig. \ref{p0837hst}) the closest objects are clearly
resolved and appear as galaxies, in particular object \#17, the object
about 7'' to the South-West, and those about 11'' to the North-East
which seem to be a close interacting pair (corresponding to object
\#20 in Table \ref{autour} and object 94 in Table I by Ellingson et
al. 1989).  The very small extension to the SE corresponds to a faint
galaxy about 1.7'' from the QSO center.

Redshift measurements are available for 7 of the 13 objects reported
as probable galaxies in Table \ref{autour}, with values between 0.0828
and 0.2677, five of them being very close to that of the quasar
(Ellingson et al. 1989).  Only one of them, object \#10 in Table
\ref{autour} (with z=0.1966), is in the field of Fig.
\ref{p0837j}. Note that in Ellingson et al. (1989) there are two
additional objects close to the QSO with measured redshifts that we
have not included in Table \ref{autour}: for that placed $\approx$ 7''
West of the QSO (88 in Ellingson et al. 1989, z=0.1929), we cannot
measure magnitudes due to contamination by the QSO; neither do we
resolve the other one, located 10'' NE of the QSO (94 in Ellingson et
al. 1989, z=0.1994).

\subsection{3C~215  (z=0.41)}

This quasar appears to have a very complex radio structure (see
e.g. Bridle et al. 1994); it is in probable tidal interaction with a
companion galaxy only 28 kpc (about 6'') away and is surrounded by 14
galaxies within $\pm 30$'' of the quasar (Hutchings \& Neff
1990). Extended ionized gas has been detected along PA=190\degr
(Crawford \& Fabian 1989). The R band profile of 3C~215.0 appears
somewhat more extended than the stellar profile, and consistent with
an elliptical host galaxy; a companion object is present 7''
South-East of the quasar, but with no measured redshift (Hutchings
1992).

In Fig. \ref{3c215j} we see a faint extension to the North-West. We
list 16 objects in Table \ref{autour}. Note that objects \#11 and \#16
have the same redshift as the quasar. Three weak objects, labeled A, B
and C are found to coincide well with three features in the Hutchings
et al. (1998) radio map.

\begin{figure}
\centerline{\psfig{figure=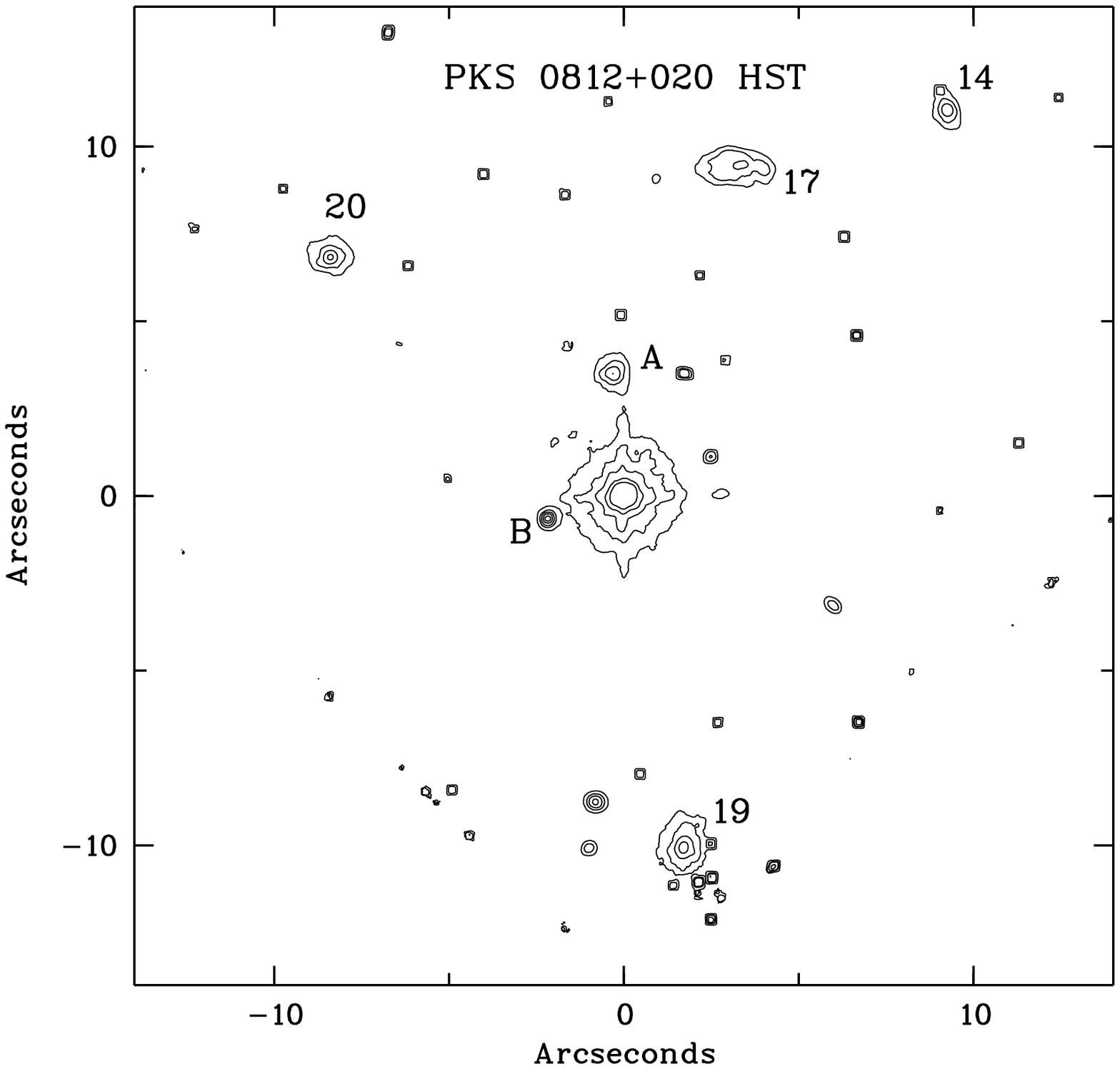,height=6cm}}
\caption[ ]{HST image of PKS~0812+020 (filter F675W). Contour levels are
6.5, 9, 15, 30 and 50 (in arbitrary units).}
\protect\label{p0812hst}
%\end{figure}
%\begin{figure}

\vspace {0.5truecm}

\centerline{\psfig{figure=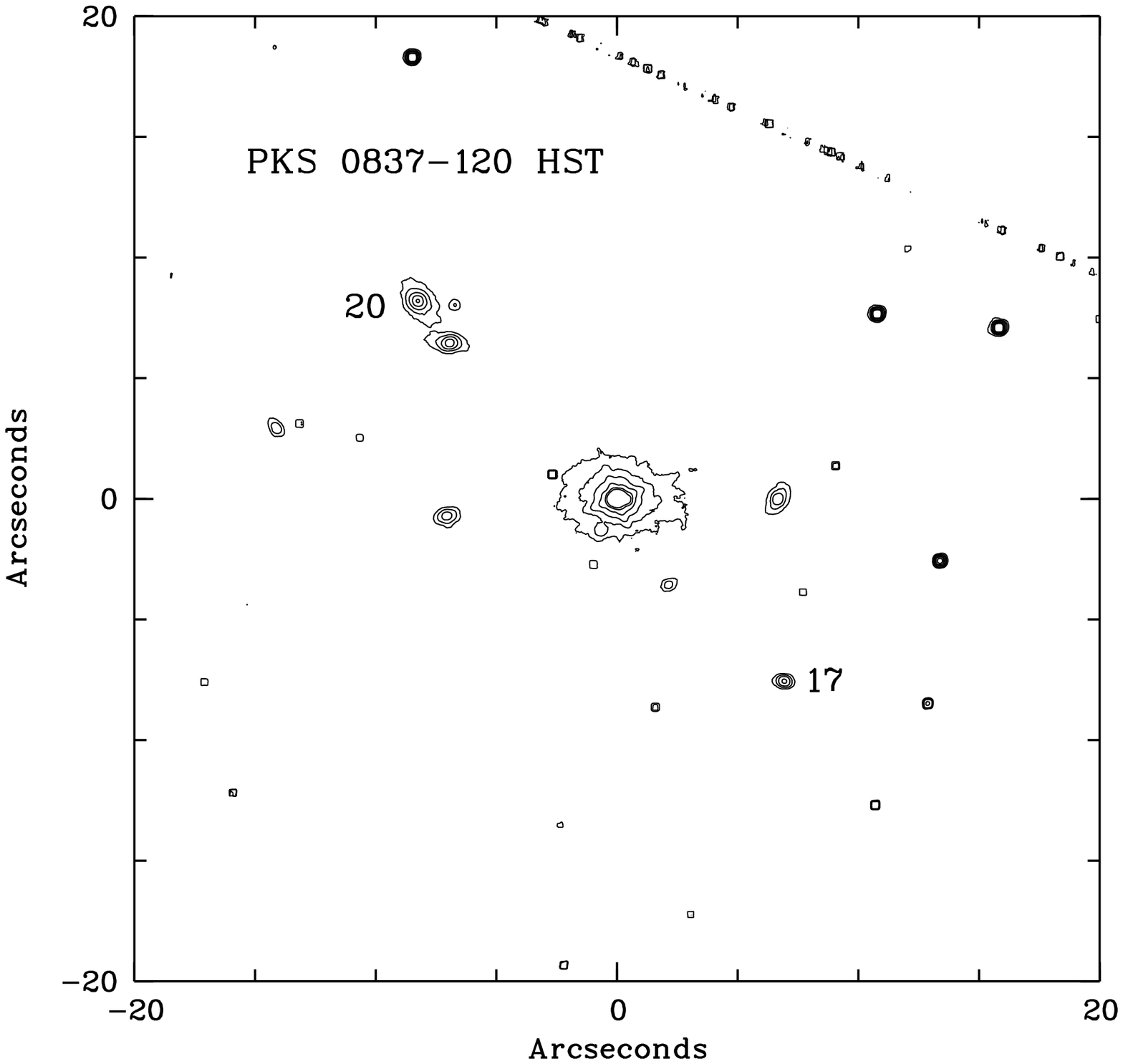,height=6cm}}
\caption[ ]{HST image of PKS~0837-120 (filter F702W). Contour levels are
2, 3, 5, 10, 20 and 30 (in arbitrary units).}
\protect\label{p0837hst}
%\end{figure}
%\begin{figure}

\vspace {0.5truecm}

\centerline{\psfig{figure=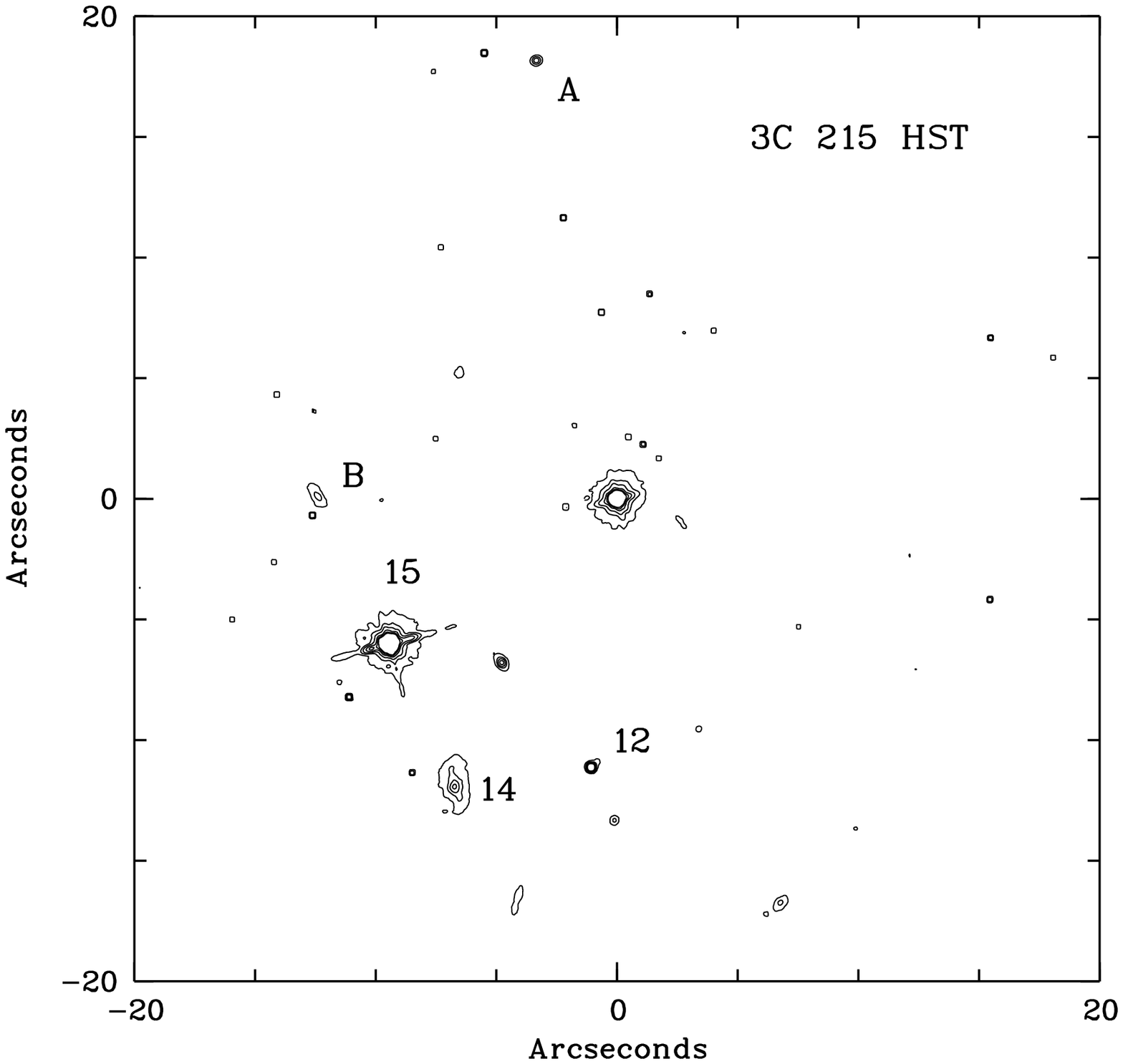,height=6cm}}
\caption[ ]{HST image of 3C~215 (filter F814W). Contour levels are
5, 7, 10, 15, 20 and 25 (in arbitrary units). }
\protect\label{3c215hst}
\end{figure}

The HST image (Fig. \ref{3c215hst}) shows that all the objects located
close to the quasar but one, that we detect in J, seem to be galaxies,
object \#14 having a spiral morphology (the brightest one, \#15, is a
star).

\subsection{IRAS~09149-6206 (z=0.057)}

The properties of the underlying galaxy have been discussed by
V\'eron-Cetty \& Woltjer (1990), who claim that a spheroidal model
gives a much better fit than a disk model to their i filter
imaging. The parameters of the fit in the V band are given in more
detail by V\'eron-Cetty et al. (1991), leading to an absolute
magnitude M$_{{\rm V}gal} = -23.0$.

In Fig. \ref{i09j} we clearly see the host galaxy of this bright
quasar, which extends about 40'' along its major axis direction
(PA$\approx$62\degr). The average isophotal profile results to be
clearly broader than that of the PSF, as shown in
Fig. \ref{profiles}. We have deconvolved the image and subtracted the
PSF to obtain a final continuous profile with no central hole (a
``flat top profile with no hole in the center'', Aretxaga et al. 1995;
Ronnback et al. 1996); the resulting profile follows quite well an
r$^{1/4}$ law; the host galaxy is therefore probably an elliptical
galaxy with magnitude K'$_{gal}$ = 10.35.  This field is very rich in
objects, but no redshift is available.

\subsection{PKS~1011-282  (z=0.253)}

The B and R band images of this quasar were only marginally resolved
by Hutchings et al. (1984, 1988). A remarkable arc-like distribution
of ionized gas was detected around this object by Stockton \& MacKenty
(1987), and analyzed by Boisson et al. (1994), in relation with its
radio properties (Gower \& Hutchings 1984).  Two galaxies are located
within $\pm 30''$ of the quasar, but there is no visible interaction
(Hutchings \& Neff 1990).
 
The extension corresponding to the host galaxy is visible in our J
image (see Fig. \ref{p1011j}) and the deconvolution enhances this
structure. However, we do not detect in J the diffuse arc-like feature
observed in the [OIII] line image (about 20'' North-West of the
quasar), implying that this is a region of ionized gas with no
underlying stellar population, as already mentioned by Boisson et
al. (1994) who detected no optical counterpart.

\subsection{3C~275.1 (z=0.557)}

Hintzen et al. (1981) have shown that this quasar could be at the
center of an elliptical galaxy belonging to a galaxy cluster. Later,
extended ionized gas with a complex structure and large scale motions
was reported by Hintzen \& Stocke (1986); the ionized gas appears to
be elongated along an angle roughly perpendicular to the radio axis
(e.g. Stocke et al. 1985). Hintzen \& Romanishin (1986) obtained
optical images in the R and redshifted narrow [OII] $\lambda$372.7
bands, and found a clear extension along PA=45\degr; once an
elliptical component has been subtracted, some small extensions remain
mainly along the major axis.  Redshifts from Ellingson \& Yee (1994)
show that the quasar has a very close companion at the same redshift
(object \#6 in Table \ref{autour}) and that it belongs to a group of
galaxies, which Kempe\'c-Krygier et al. (1998) have analyzed in detail
from the physical and dynamical points of view; they have shown that
the quasar is located at the bottom of the gravitational potential
well of the group.

Our J band image (Fig. \ref{3c275j}) shows the host galaxy as
elliptical contours with a PA roughly consistent with that found in
the optical images. There is also a small extension to the
East. Deconvolution is not possible since the stars in the frame are
too faint or contaminated by close objects to give an accurate
estimation of the PSF. Several nearby objects are also visible in the
J image, a number of these being foreground galaxies (see redshifts in
Table \ref{autour}).

The HST image (Fig. \ref{3c275hst}) shows that object \#4 has a spiral
morphology and that objects \#9 and \#6 are galaxies, \#6 being at the
same redshift as the quasar. Besides, we can note that galaxies \#4
and \#6 correspond to two small radio emitting features in the
Hutchings et al. (1998) radio map.

Note that Akujor et al. (1994) suggest that the northern radio
component may be distorted by interaction with a companion galaxy
(object \#4). This is unlikely since the redshift of this galaxy is
notably smaller than that of the quasar.

\subsection{PKS~1302-102   (z=0.286)}

Hutchings \& Neff (1992) have detected two objects in the close
vicinity of the quasar and related their presence to the merging state
of the system; they also fit a profile which is a combination of an
r$^{1/4}$ law and an exponential law.  Bahcall et al. (1995) did not
detect the host galaxy in their HST image.  After deconvolution with
the PSF however they can see two companion galaxies located about 1''
and 2'' N-NW away from the quasar center (see their Figure 8). From a
new HST image, Disney et al. (1995) detect the two companion objects
with no need of PSF subtraction and the host galaxy is well fit 
with a r$^{1/4}$ profile. The detection of the host galaxy from HST
imaging was later confirmed by Bahcall et al. (1997), who gave values
for the size and morphology of the host galaxy.

Our J band image (Fig. \ref{p1302j}), very similar to that of McLeod
\& Rieke (1994b) in the H band, is well suited to trace the host
galaxy; its isophotal profile is clearly different from that of the
PSF (see Fig. \ref{profiles}) as in McLeod \& Rieke (1994b), and
notably more extended than that obtained by Wyckoff et al. (1981). The
underlying galaxy is extended along a direction roughly perpendicular
to the radio axis, as already noted by Gower \& Hutchings (1984). In
the deconvolved image we have subtracted the quasar by using the PSF
normalized to a central value that produces no hole after
subtraction. This barely allows us to detect the object 2'' North of
the quasar, which appears as an isophotal distorsion in the direct
image, but we do not separate the quasar from the small object 1''
away. The contribution of the quasar to the total light in J results
to be about 60\%. The resulting profile of the host galaxy can be well
fit by an r$^{1/4}$ law and its derived magnitude is J$_{gal}$ = 14.9. These
results are in agreement with McLeod \& Rieke (1994b) concerning the
relative contribution of the host galaxy (31\% with H$_{gal}$ = 14.79)
although they fit an exponential. We note that the presence of the
very close object 1'' away does not produce any significant
contamination in the average profile, which is dominated by the host
galaxy.

\subsection{3C~281 (z=0.599)}

This quasar is a double radio source (Hutchings et al. 1998) rich in
extended ionized gas (Bremer et al. 1992).

In our J band image (Fig. \ref{3c281j}) an object 5'' South of the
quasar is visible, together with extensions to the North and to the
East. The deconvolution enhances the Northern extension and separates
the Eastern one as an object.

The HST image (Fig. \ref{3c281hst}) shows that these extensions are in
fact two objects located respectively 3.5'' North and 2.7'' East of
the quasar center. We also see that objects \#6 and \#9 in Table
\ref{autour}, which were reported as a probable galaxy and a star
respectively, are both galaxies. Object 132 in Yee et al. (1986),
located 5'' North of \#9 is also a galaxy, but is too weak to be
detected in the J band image.

\subsection{4C~20.33 (z=0.871)}

This quasar is an asymmetric radio source with a collimated one-sided
jet extending from the quasar to the South (Mantovani et al. 1997).
Fabian et al. (1988) detected extended ionized gas around this quasar.

Isophotes are elongated towards the South in our J band image
(Fig. \ref{4c20j}), a feature which is enhanced after
deconvolution. In this case, and contrary to some of the objects
previously described, the elongation appears to be along a direction
roughly similar to that of the radio axis (Mantovani et al. 1992). We
also see a number of objets (probably galaxies) close to the quasar
that could be associated with it. Redshifts are obviously needed for
this field.

\subsection{4C~11.50 (z=0.436)}

The image by Stockton and MacKenty (1987) shows extended ionized gas
elongated along PA $\approx$ 306\degr. A second quasar (object \#8) at
much higher redshift (z=1.901, Wampler et al. 1973) is located 5'' to
the East and a tight group of three galaxies (object \#7) is
associated with 4C~11.50 at an average distance of 10'' West of the
quasar (Stockton 1978).

\begin{figure}
\centerline{\psfig{figure=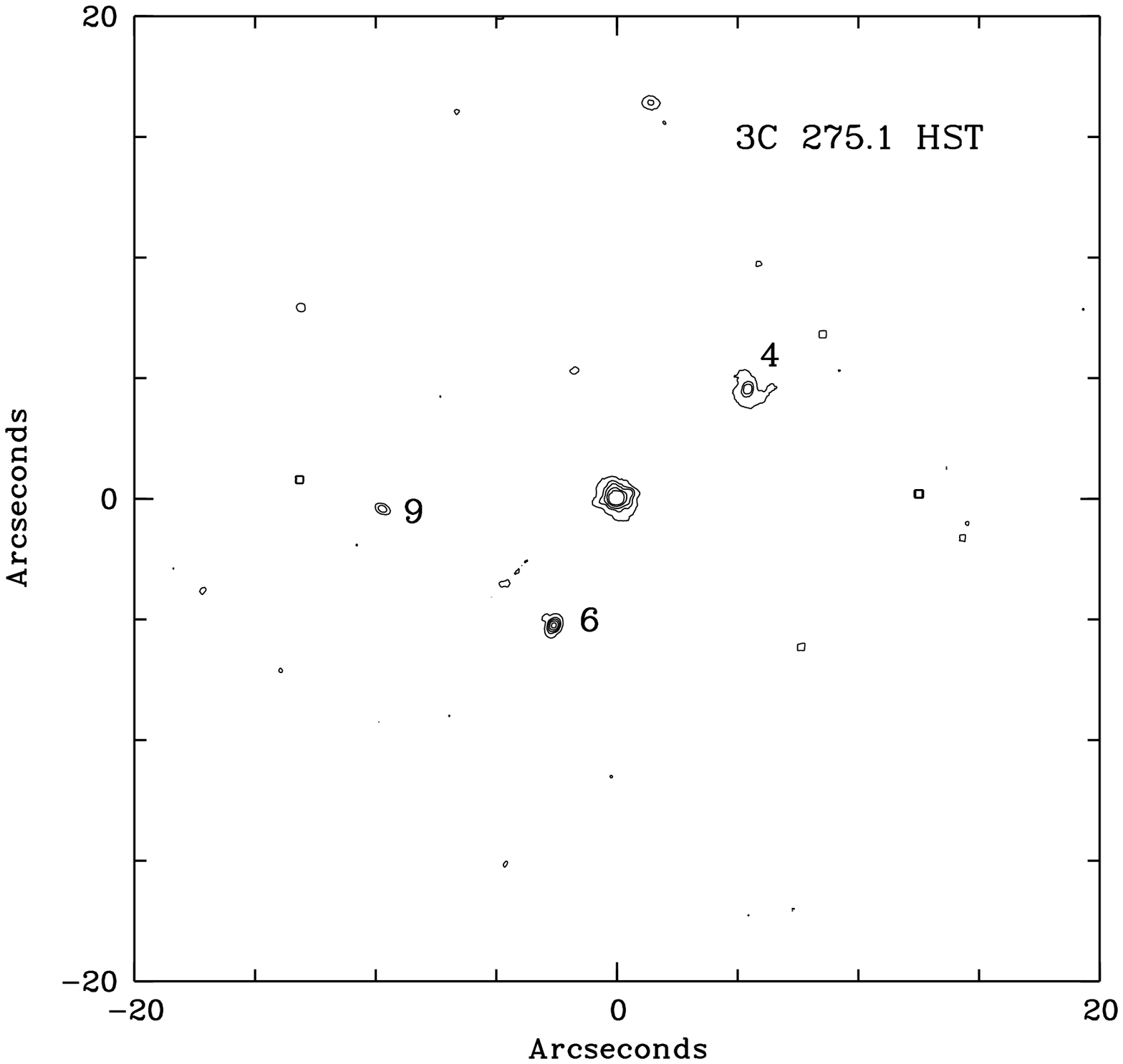,height=6cm}}
\caption[ ]{HST image of 3C~275.1 (filter F675W). Contour levels are
2.9, 4, 5, 7 and 10 (in arbitrary units).}
\protect\label{3c275hst}
%\end{figure}
%\begin{figure}

\vspace {0.5truecm}

\centerline{\psfig{figure=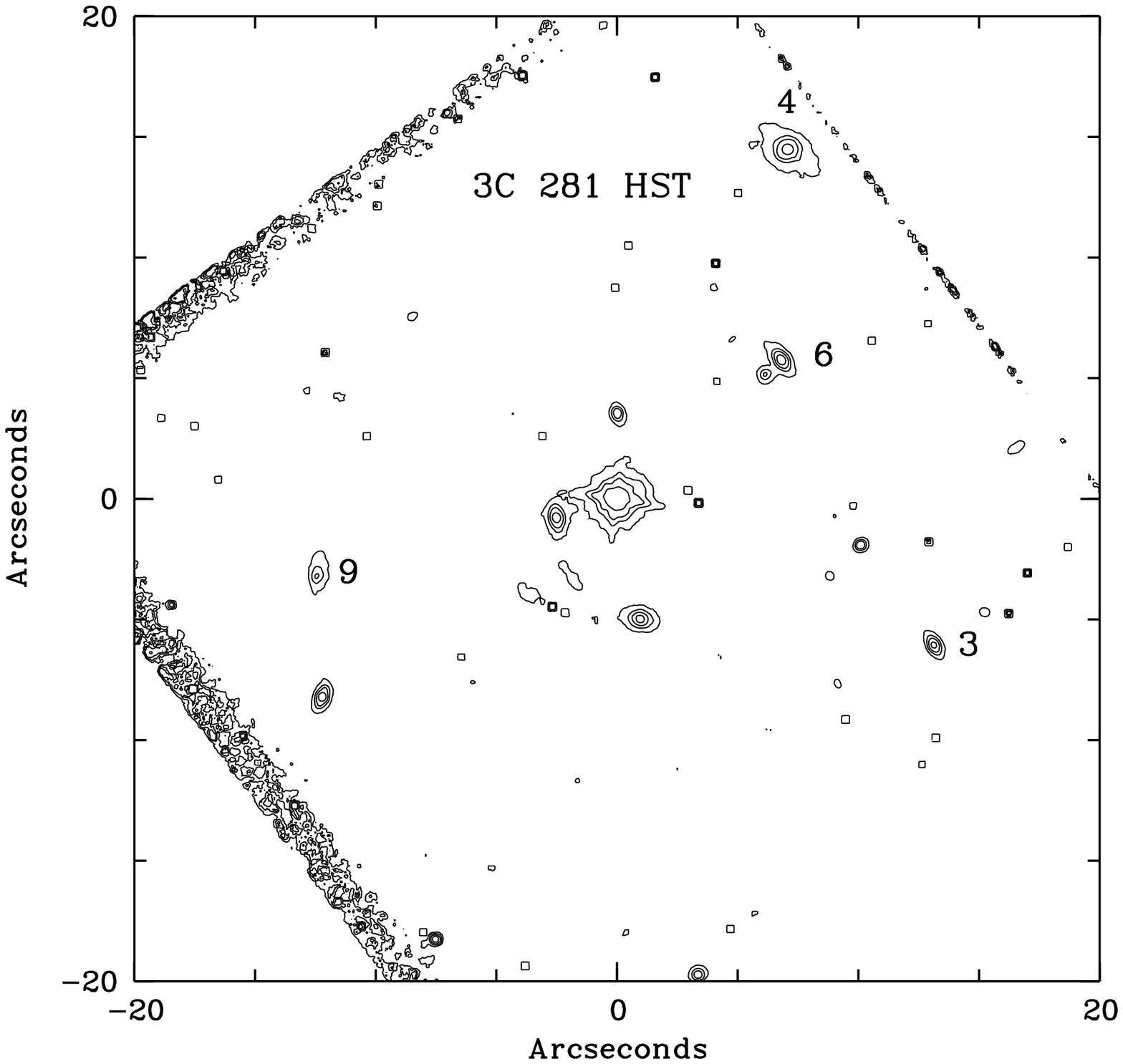,height=6cm}}
\caption[ ]{HST image of 3C~281 (filter F814W). Contour levels are
3.5, 5, 7 and 15 (in arbitrary units).}
\protect\label{3c281hst}
%\end{figure}
%\begin{figure}

\vspace {0.5truecm}

\centerline{\psfig{figure=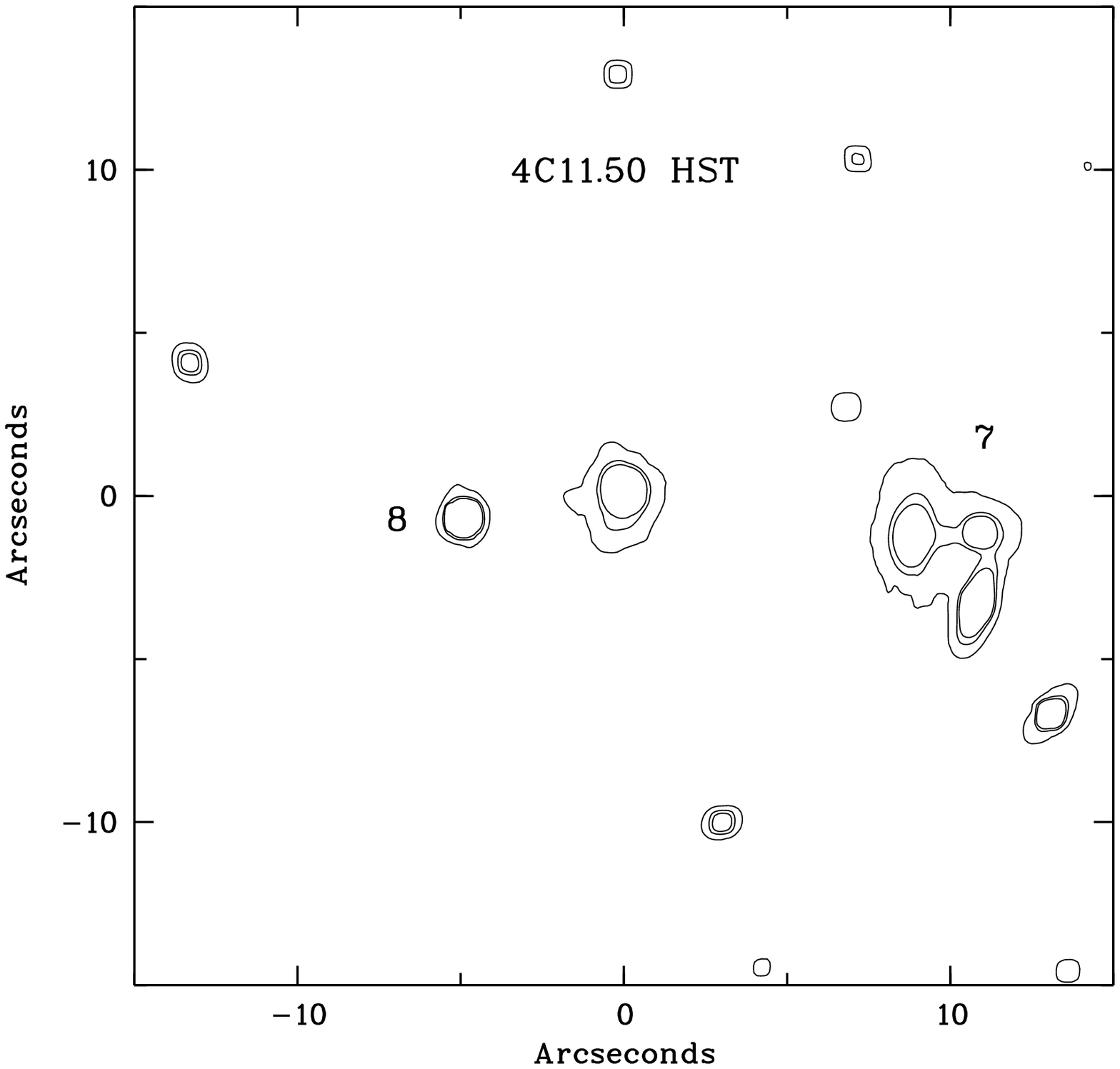,height=6cm}}
\caption[ ]{HST image of 4C~11.50 (filter F702W). Contour levels are
13, 20 and 25 (in arbitrary units).}
\protect\label{4c1150hst}
\end{figure}

We observe similar structures in our J and K' band images
(Figs. \ref{4c1150j}, \ref{4c1150k}).  The quasar appears elongated
along PA $\approx$ 47\degr~ in our deconvolved K' image, where the
three objects to the West reported by Stockton and Mac\-Kenty (1987) are
clearly separated.  The northernmost one is elongated to the West, due
to the presence of a separate object as seen in the HST image
(Fig. \ref{4c1150hst}). The object detected 12'' South-West of the
quasar (object 142 in Yee et al. 1986) is extended in the HST image,
but is too weak to be detected in J.  Note that the host galaxy of the
quasar appears to be detected in the HST image.  A weak object labeled
A appears to coincide with a faint feature in the Hutchings et
al. (1998) radio map.

We give R magnitudes from the literature for the objects in the field
in Table \ref{autour}; note that the quasar at z=1.901 (object \#8) is
indicated as a star. We also give redshifts for objects \#7 and \#9
from Ellingson et al. (1991a); both objects result to be companion
galaxies to the quasar (z=0.4323 and 0.4331 respectively).
 
\subsection{Mrk~877 (z=0.114)}

Boroson et al. (1982) detected faint emission from ionized gas
extended over several arcseconds from Mrk~877. The analysis of the H
band image by McLeod and Rieke (1994a) results in an quasar isophotal
profile which is slightly different from that of the PSF.  Our J band
image (Fig. \ref{mk877j}) has been taken in poor seeing conditions
(1.4 arcsec) but good spatial sampling (0.5 arcsec/pix) and shows a
definite difference between the QSO and PSF profiles (see
Fig. \ref{profiles}).  The profile from our deconvolved J band image
can be well fit by an $r^{1/4}$ law, suggesting here also that the
host galaxy to this quasar is an elliptical galaxy, with an estimated
magnitude J$_{gal}$=14.95. A number of objects appear in the QSO field
with profiles different from the PSF and sizes comparable to the QSO
host and could be possible companion galaxies.

\subsection{3C~334 (z=0.555)}

This quasar is a triple radio source oriented NW to SE with a jet
originating in the nucleus and extending towards the SE component
(Hintzen et al. 1983, Swarup et al. 1984, Dennet-Thorpe et al. 1997,
Hutchings et al. 1998); it has a resolved underlying galaxy (Hintzen
1984). Ionized gas was found to extend over 6'' in the spectra by
Crawford \& Fabian (1989), and an object was detected 7'' south of the
quasar by Hes et al. (1996) using narrow band imaging in the [OII]
$\lambda$372.7 line.

\begin{figure}
\centerline{\psfig{figure=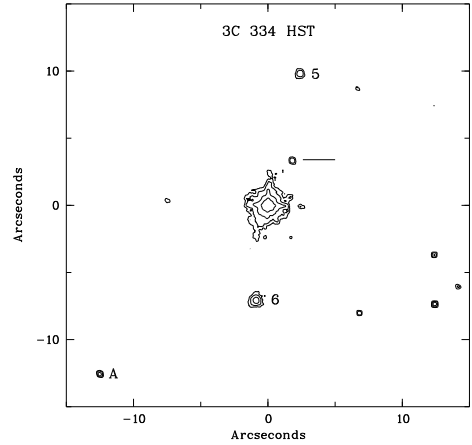,height=6cm}}
\caption[ ]{HST image of 3C~334 (filter F675W).Contour levels are
0.8, 1, 2 and 7 (in arbitrary units).}
\protect\label{3c334hst}
\end{figure}

Our J band image (Fig. \ref{3c334j}) shows an extension to the
North-West (to about 5'' from the quasar) and two objects 10'' North
and 7'' South respectively (objects \#5 and \#6 in Table
\ref{autour}).  The latter corresponds to that detected in [OII] by
Hes et al. (1996). A weak object, labeled A, coincides with a bright
feature in the Hutchings et al. (1998) radio map and is also visible
in the HST image (Fig. \ref{3c334hst}).

The deconvolved image allows to separate the small extension to the
North-West as an object (indicated in Fig. \ref{3c334hst} with a
line). This object, as well as objects \#5 and \#6, are clearly
separated from the quasar in the HST image, \#6 probably being a
galaxy.  It would be very interesting to determine the redshift of
these three objects and to search for any relation with the three weak
Ly$\alpha$ absorptions observed at $z$~=~0.5387, 0.5449 and 0.5491
(Jannuzi et al. 1998).

\subsection{MC~1745+163    (z=0.392)}

Stockton and MacKenty (1987) give an image of the ionized gas in
MC~1745+163, which they claim to be resolved, with an elongation to
the West along PA $\approx$ 90\degr.

\begin{figure}
\centerline{\psfig{figure=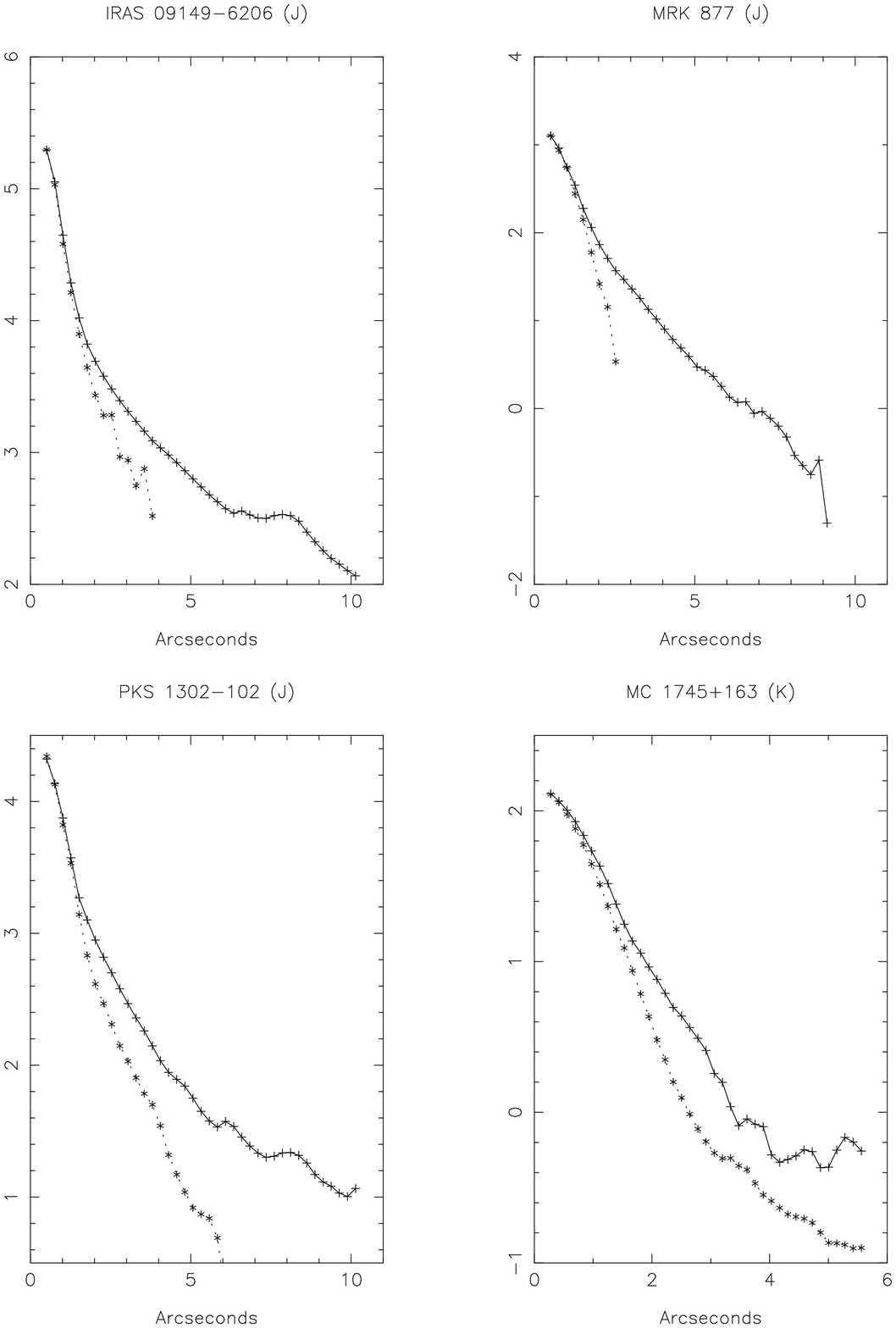,height=12cm}}
\caption[ ]{Isophotal profiles of quasars (crosses) that 
result to be clearly different from those of the corresponding PSFs (stars) 
which have been normalized to the QSO for direct comparison. Fluxes are in 
arbitrary units.}
\protect\label{profiles}
\end{figure}

Our J band image shows this extension as well; though in the direct
image (Fig. \ref{m1745j}) the major axis PA is smaller (PA =50\degr ),
it is somewhat larger (62\degr ) in the deconvolved image. The
isophotal profile is notably different from that of the PSF already in
the direct image (Fig. \ref{profiles}). The resulting profile of the
deconvolved and PSF subtracted image is well fit by an $r^{1/4}$
profile; the host galaxy is therefore likely to be an elliptical with
K'$_{gal}$ = 16.19. Many objects are detected in this field; most are
stellar like, except for objects \#12 and \#16 which look like
galaxies but for which we do not have redshifts.

\subsection{4C~11.72 = PKS 2251+113 (z=0.323)}

This quasar has two companion galaxies with comparable redshifts: 
$z$~=~0.3287 et 0.3240 for objects \#4 and \#5 in Table \ref{autour}
respectively (Robinson \& Wampler 1972, following observations by Gunn
1971). It is embedded in an extended ionized nebulosity (Stockton \&
MacKenty 1987, Hutchings \& Crampton 1990), where the gas appears to
be very disturbed, oppositely to the host galaxy, which Hutchings \&
Neff (1992) classify as an ``undisturbed $r^{1/4}$ galaxy''; from
integral field spectroscopy, Durret et al. (1994) have shown that the
velocity field was in fact consistent with the presence of a rotating
disk, on to which are superimposed several smaller blobs possibly
interacting with the main envelope.

Optical images in $b$ and $v$ by Block and Stockton (1991) show the
extended nature of the residual after the QSO subtraction (see their
Figure 5). Our J image is given in Fig. \ref{4c1172j}. Since we do
not have in our field a suitable star to compute the PSF, we cannot
proceed either with the deconvolution nor with a PSF subtraction to
look for the nearest objects, labeled A and B by Block \& Stockton
(1991).  Objects \#9 and \#10 in Table \ref{autour} correspond to
objects 7 and 4 in Block \& Stockton (1991). Object \#7 corresponds to
two objects, the northernmost one corresponding to object W in Gunn
(1971).

Jannuzi et al. (1998) have observed this quasar spectroscopically with
the HST and detected a strong C~{\sc iv}-O~{\sc vi} associated
absorption line system at $z$~=~0.3256~$\sim$~$z_{\rm em}$.  It would
be very interesting to obtain higher spectral resolution optical data
to study the kinematics and composition of the gas, in order to
investigate the possibility that it is associated either with the AGN
(Petitjean et al. 1994) or with a hot phase resulting from the
interaction of several objects (Durret et al. 1994).

There is an additional Ly$\alpha$ absorption line at $z$~=~0.3236. If
the latter is associated with the gaseous halo of object \#5 which has
the same redshift, then the radius of the halo should be larger than
95$h^{-1}_{50}$~kpc for $q_{\rm o}$~=~0. This radius is consistent
with studies of the low-redshift Ly$\alpha$ forest (e.g. Le~Brun et
al. 1996).  However this quasar is part of a group of galaxies where
the distribution of gas into individual halos is questionable.  Indeed
there is no H~{\sc i} absorption associated with object \#4 which is
at a projected distance of 135$h^{-1}_{50}$~kpc from the line of
sight.


\section{Summary and Conclusions}

We have obtained near infrared imaging for a sample of fifteen QSOs,
most of them selected as having extended ionized nebulosities, in
order to analyze the possible connection of this gas with the
underlying stellar population. We detect infrared extensions in at
least eleven of them. The ionized gas extensions obtained from
spectroscopic studies are within the extensions that we determine from
NIR imaging. When images of the ionized gas are available, their
extensions are compatible with the gas belonging to the host galaxies.
The kinematical behaviour of this gas should be analyzed in order to
be able to put some constraints on the possible mechanisms for the gas
origin and fuelling.

We have determined the relative contribution of the host galaxies by
subtracting the scaled PSF to the quasar profile. The corresponding
magnitudes are given in Table 3. For the whole set of 15 objects, the
resulting average contribution is 50 $\pm$ 20 \% in J (48 $\pm$ 22 \%
for the twelve radio loud quasars and 58 $\pm$ 7 \% for the three
radio quiet). We could extract the host galaxy for only four quasars
in K', for which the host galaxy contributed 67 $\pm$ 8 \% in average
to the total luminosity.  For the four hosts extracted in both J and
K', we find that their contributions are similar in both bands for
IRAS~09149-6206 (radio quiet) and larger in K' than in J for the
remaining three (radio loud), in agreement with the expectation that
K' describes the stellar population better than J. The average
absolute magnitudes of the host galaxies are $M_{\rm J} = -25.5
\pm$1.1 ($-25.7 \pm$1.2 for the radio loud and $-24.9 \pm$0.5 for the
radio quiet) and $M_{\rm K'} = -27.0 \pm$0.6.

This is in agrement with previous results on radio loud quasars in
other bands (V\'eron-Cetty \& Woltjer 1990, Taylor et al. 1996,
Bahcall et al. 1997, Hooper et al. 1997, assuming typical values
V$-$H=3.0, R$-$H=2.5, H$-$K=0.2 and J$-$H=0.8). J and K' absolute
magnitudes are also compatible with those of Bright Cluster Member
galaxies by Thuan \& Puschell (1989) and Arag\'on-Salamanca et
al. (1998).

Given our limited spatial resolution, we cannot always discriminate
whether asymmetric extensions are intrinsic or are the consequence of
the presence of objects in projection close to the line of sight to
the quasar. We have attempted to answer this question first by
estimating the shape of the PSF from starlike objects in the field,
deconvolving the images and then subtracting the PSF of the
deconvolved image to the quasar infrared profile, and second by using
high resolution optical images retrieved from the HST archives, which
were available for eight of the observed QSOs.

It was possible to resolve the extensions into close objects for
PKS~0812+020, PKS~0837-120 and 3C 281, though for the second quasar
the elongated structure seems also to be present in the HST images,
suggesting that the underlying galaxy is indeed detected.  In all the
other cases the infrared extensions are symmetric, as expected for a
normal underlying host galaxy; however, the corresponding average
profiles are notably different from the PSF profiles in only four
QSOs: IRAS~09149-6206, PKS~1302-102, Mrk~877 and MC~1745+163 (see
Fig.\ref{profiles}), where the harboring galaxies are well fit by the
r$^{1/4}$ profile describing elliptical galaxies. Out of these four
quasars, note that PKS~1302-102 and MC~1745+163 are radio loud, while
IRAS~09149-6206 and Mrk~877 are radio quiet.

Note that although HST images were needed to resolve close objects,
they have proven to be sometimes less efficient to detect low surface
brightness features, as for PKS~1302-102 (Bahcall et al. 1995, see
section 3.8).  For 3C~215, the HST image does not allow to trace the
elongation due to the host galaxy, whereas our J image does. The
present study confirms that infrared imaging is well adapted to detect
underlying galaxies in quasars, after objects located close to the
quasar are identified with high spatial resolution imaging (HST or
adaptive optics ground based data).

One of the possibilities to explain the presence of extended gaseous
envelopes around QSOs is to claim their belonging to rich
environments, with the presence of close companions producing the
physical mechanisms (tidal forces or nonaxisymmetrical perturbations
in accretion or merger events) that can account for gas fuelling.  A
number of redshifts have been measured for the objects in the vicinity
of five of the QSOs we have observed: PKS~0812+020, PKS~0837-120, 3C
215, 3C 275.1 and 4C 11.50. Note that out of these five quasars four
are known to be rich in ionized gas, and the only one for which
ionized gas has not been reported is PKS~0837-120, which belongs to a
richness class 1 cluster. For all five quasars, between one and five
galaxies were found to have redshifts similar to that of the quasar.
The presence of such a large number of companion objects is therefore
likely to imply the existence of environmental effects which could
account for the existence of extended ionized nebulosities in most of
these objects.

Unfortunately, we do not have redshift informations for field objects
around the ten other QSOs in our sample. However, our infrared imaging
shows that at least in projection most of these quasars have a number
of galaxies in their environments.  3C 334 has at least one close
galaxy, and some of the objects close to PKS~1011-282 and 4C 20.33 seem
to be galaxies as well, while the fields around IRAS 09149-6206 and
MC~1745+163 are crowded with mostly star like objets.

In order to improve such a study of the physical parameters of the
underlying galaxies of quasars, and to describe their environments, high
spatial resolution deeper infrared images of larger fields and with a
higher sensitivity are obviously required (in particular, our limiting
magnitudes together with the small frame sizes we are dealing with --
less than 1'~-- result in small object numbers that prevent a reliable
statistical analysis of colors and number counts of companions, such
as that performed by Hall \& Green 1998). After the neighbouring
objects are detected and galaxies are separated from stars,
spectroscopy is necessary to determine which galaxies really belong to
the quasar environment.

\begin{acknowledgements}

I.~M\'arquez acknowledges financial support from the Spanish
Ministerio de Educaci\'on y Ciencia.  I. M. acknowledges technical
support from the IRAC team at ESO, specially Luis Ramirez. We are very
grateful to Ron Probst, who made available the SQIID package for the
reduction of infrared images within IRAF available to us. Finally, we
thank the anonymous referee for useful comments.

\end{acknowledgements}

\voffset 0.0truecm
\begin{table*}
\caption{Magnitudes and positions for quasars and nearby detected objets}

\begin{tabular}{rrrrrrr|rrrr}
\hline
object & $\alpha$(1950) & $\delta$(1950) & $\Delta \alpha$ & $\Delta \delta$ & $J$ & $K'$ & $R$ & ref$^a$ & nr. & type($z$)$^b$ \\
\hline
A~0401-350A & 04 01 01.20&-35 03 40.0&    0.0  &   0.0 & 15.5& &&&&QSO \\
	  1 & 04 00 58.48&-35 02 38.4&   43.0  & -16.8 & 18.7& &&&&3\\
	  2 & 04 00 59.42&-35 02 51.8&    6.5  &   2.9 & 19.5& &&&&1\\
	  3 & 04 01 00.13&-35 02 56.3&   65.9  &  18.7 & 16.8& &&&&2\\
	  4 & 04 01 01.73&-35 03 37.1&  -13.1  &  43.7 & 18.5& &&&&2\\
	  5 & 04 01 04.70&-35 03 56.8&  -21.9  &  48.2 & 18.4& &&&&3\\
	  6 & 04 01 04.71&-35 02 51.5&   43.0  &  48.5 & 12.1& &&&&3\\
	  7 & 04 01 06.57&-35 03 21.3&  -33.4  &  61.6 & 11.7& &&&&3\\
\hline
PKS 0812+020 & 08 12  47.26& 02 04 13.1& 0.0   & 0.0    & 15.6 & 14.5 & & 1  &  & QSO \\
	  1 & 08 12  43.45& 02 05 05.2& -57.1 &   52.1 & 14.4 &      & & & & 3\\
	  2 & 08 12  43.63& 02 03 27.8& -54.4 &  -45.3 & 14.7 &      & & & & 3\\
	  3 & 08 12  43.65& 02 04 12.5& -54.2 &   -0.6 & 18.6 &      & & & & 3\\
	  4 & 08 12  43.80& 02 04 44.9& -51.8 &   31.8 & 17.5 &      & 19.0 & 1  & 3 & 3 \\
	  5 & 08 12  43.91& 02 03 35.3& -50.2 &  -37.8 & 15.6 &      & & & & 3\\
	  6 & 08 12  44.01& 02 04 04.4& -48.8 &   -8.7 & 17.2 &      & 18.5 & 1  & 4 & 1(0.3478) \\
	  7 & 08 12  44.15& 02 04 59.7& -46.6 &   46.6 & 16.6 &      & & & & 3\\
	  8 & 08 12  44.61& 02 03 44.7& -39.7 &  -28.4 & 17.6 &      & 20.0 & 1  & 5 & 3 \\
	  9 & 08 12  44.89& 02 04 28.7& -35.6 &   15.6 & 16.3 &      & & & & 3\\
	 10 & 08 12  45.10& 02 03 16.9& -32.4 &  -56.2 & 18.1 &      & & & & 3\\
	 11 & 08 12  45.38& 02 03 45.1& -28.3 &  -28.0 & 17.8 & 17.3 & 20.0 & 1  & 6 & 1(0.30)\\ 
	 12 & 08 12  45.40& 02 03 21.9& -28.0 &  -51.2 & 18.6 &      & & & & 3\\
	 13 & 08 12  45.57& 02 04 45.4& -25.3 &   32.3 & 17.7 &      & & & & 3\\
	 14 & 08 12  46.65& 02 04 24.2&  -9.1 &   11.1 & 19.0 & 18.0 & & & & 3\\
	 15 & 08 12  46.98& 02 04 58.4&  -4.2 &   45.3 & 15.2 &      & & & & 3\\
	 16 & 08 12  47.00& 02 04 41.7&  -3.9 &   28.6 & 16.3 & 15.9 & & & & 3\\
	 17 & 08 12  47.08& 02 04 22.6&  -2.7 &    9.5 & 18.8 & 17.7 & 20.3 & 1 & 9 & 2\\
	 18 & 08 12  47.12& 02 03 14.1&  -2.0 &  -59.0 & 15.6 &      & & & & 3\\
	 19 & 08 12  47.14& 02 04 03.0&  -1.8 &  -10.1 & 17.6 & 16.6 & 20.2 & 1 &10  & 1(0.4030)\\ 
	 20 & 08 12  47.82& 02 04 20.1&   8.4 &    7.0 & 18.5 & 17.4 & & & & 3\\
	 21 & 08 12  48.47& 02 04 55.0&  18.0 &   41.9 & 18.7 &      & & & & 3\\
	 22 & 08 12  48.54& 02 05 04.7&  19.2 &   51.6 & 17.7 &      & 20.0 & 1  & 12 & 1(0.4038)\\
	 23 & 08 12  48.97& 02 05 12.6&  25.6 &   59.5 & 17.4 &      & & & & 3\\
	 24 & 08 12  49.56& 02 04 36.0&  34.5 &   22.9 & 18.6 &      & 19.7 & 1 & 14 & 3 \\
	 25 & 08 12  50.04& 02 04 19.0&  41.6 &    5.9 & 17.7 & 16.7 & 20.2 & 1  & 15 & 1(0.3599) \\
	 26 & 08 12  50.20& 02 04 19.7&  44.1 &    6.6 & 17.6 & 17.4 & & & & 2\\
	 27 & 08 12  51.33& 02 03 41.9&  61.0 &  -31.2 & 18.1 &      & & & & 3\\
	 28 & 08 12  51.63& 02 04 17.5&  65.5 &    4.4 & 16.8 &      & & & & 3\\
\hline
PKS 0837-120 & 08 37  27.95&-12 03 54.2&    0.0 &    0.0 & 15.0 & 13.4  & 16.4 & 2 & 90 &QSO\\
	  1 & 08 37  23.49&-12 03 06.7&  -65.4 &   47.5 & 14.6 &       & 16.2 & 2 & 54 & 3\\
	  2 & 08 37  23.59&-12 04 21.9&  -63.9 &  -27.7 & 14.9 &       & 16.2 & 2 & 55 & 3\\
	  3 & 08 37  23.66&-12 04 29.3&  -63.0 &  -35.1 & 18.5 &       & 20.4 & 2 & 56 & 2\\
	  4 & 08 37  24.15&-12 04 30.1&  -55.7 &  -35.9 & 19.0 &       & 20.4 & 2 & 59 & 3\\
	  5 & 08 37  24.34&-12 03 19.3&  -53.0 &   34.9 & 17.0 &       & 18.9 & 2 & 61 & 1(0.2677)\\
	  6 & 08 37  24.38&-12 03 39.7&  -52.3 &   14.5 & 14.9 &       & 16.0 & 2 & 64 & 3\\
	  7 & 08 37  24.65&-12 02 00.1&  -48.3 &   54.1 & 18.8 &       & 20.5 & 2 & 65 & 3\\
	  8 & 08 37  25.19&-12 03 51.8&  -40.5 &    2.4 & 16.8 & 16.0  & 20.1 & 2 & 67 & 3\\
	  9 & 08 37  25.41&-12 02 43.0&  -37.3 &   71.2 & 15.9 &       & 17.3 & 2 & 69 & 3\\
	 10 & 08 37  26.03&-12 03 54.8&  -28.1 &   -0.6 & 17.6 & 16.5  & 19.7 & 2 & 76 & 3(0.1966)\\
	 11 & 08 37  26.39&-12 04 08.4&  -22.9 &  -14.2 & 18.4 & 17.1  & 20.2 & 2 & 79 & 3\\
	 12 & 08 37  26.69&-12 04 00.7&  -18.5 &   -6.5 & 18.3 & 17.1  & 20.4 & 2 & 81 & 3\\ 
	 13 & 08 37  26.89&-12 03 46.5&  -15.6 &    7.7 & 17.2 & 16.8  & 20.3 & 2 & 83 & 3\\
	 14 & 08 37  26.93&-12 04 48.4&  -15.0 &  -54.2 & 18.1 &       &      &   &    & 2\\
	 15 & 08 37  27.01&-12 03 02.7&  -13.7 &   -2.6 & 19.4 &       & 20.8 & 2 & 85 & 3\\
	 16 & 08 37  27.02&-12 04 56.8&  -13.7 &   -8.5 & 20.1 &       & 2.44 & 2 & 86 & 3\\
	 17 & 08 37  27.47&-12 04 01.8&   -7.1 &   -7.6 & 20.1 & 17.7  & 21.6 & 2 & 89 & 1\\
	 18 & 08 37  28.08&-12 02 54.6&    2.7 &   60.2 & 16.3 &       & 17.4 & 2 & 91 & 3\\
	 19 & 08 37  28.14&-12 04 54.0&    1.9 &  -60.4 & 19.0 &       & 23.7 & 2 & 92 & 0$^c$\\
	 20 & 08 37  28.49&-12 03 46.2&    8.0 &    8.0 & 17.7 & 17.1  & 20.2 & 2 & 96 & 2\\
	 21 & 08 37  28.53&-12 03 35.8&    8.5 &   18.4 & 19.3 &       & 20.2 & 2 & 95 & 3\\
	 22 & 08 37  29.45&-12 03 43.3&   22.0 &   10.9 & 19.1 & 18.1  & 20.5 & 2 & 101& 3\\
	 23 & 08 37  29.55&-12 03 33.9&   23.4 &   20.3 & 18.6 & 17.2  & 20.5 & 2 & 102& 1\\
	 24 & 08 37  30.18&-12 03 49.6&   32.6 &    4.6 & 16.4 & 16.0  & 17.6 & 2 & 107& 3\\
\hline
\end{tabular}
%\end{center}

\begin{footnotesize}

$^a$ 1: Ellingson et al. (1991a); 2: Ellingson et al. (1989); 
3: Ellingson et al. (1991b); 4: Yee et al. (1986); 5: Hintzen (1984);
6: Green \& Yee (1984); 7: Ellingson et al. (1994); 8: Kirhakos et al. (1994); 
9: Ellingson \& Yee (1994); 10: Robinson \& Wampler (1972).

$^b$ Types are 1=galaxy, 2=probable galaxy, 3=star. In parentheses we give 
either available redshifts or other references for photometry in the form 
(reference, object number in that reference, z when available)

$^c$ Cosmic ray or noise event in ref. 2. 
\end{footnotesize}
\protect\label{autour}
\end{table*}

\newpage

\begin{table*}[h]
{\bf Table 2.} (Cont.)

%\begin{center}
\begin{tabular}{rrrrrrr|rrrr}
\hline
object & $\alpha$(1950) & $\delta$(1950) & $\Delta \alpha$ & $\Delta \delta$ & $J$ & $K'$ & $R$ & ref & nr. & type($z$) \\
\hline
	 25 & 08 37 30.96&-12 03  14.9 &   44.1 &   39.3 & 16.1 &       & 17.4 & 2 &114 & 3\\
	 26 & 08 37 31.18&-12 04  20.9 &   47.4 &  -26.7 & 19.0 &       & 21.3 & 2 &116 & 3\\ 
	 27 & 08 37 31.31&-12 04  09.3 &   49.3 &  -15.1 & 14.3 &       & 15.3 & 2 &117 & 3\\
	 28 & 08 37 31.35&-12 03  43.2 &   49.9 &   11.0 & 11.9 &       & 13.9 & 2 &120 & 3\\
	 29 & 08 37 31.35&-12 04  48.0 &   49.9 &  -53.8 & 17.8 &       &      & 2 &118 & 2(0.1969)\\ 
	 30 & 08 37 31.43&-12 03  09.3 &   51.0 &   44.9 & 18.1 &       & 19.7 & 2 &119 & 1\\
	 31 & 08 37 31.44&-12 05  04.0 &   51.1 &  -69.8 & 17.8 &       & 19.6 & 2 &122 & 1(0.1994)\\
	 32 & 08 37 31.70&-12 03  18.0 &   54.9 &   36.2 & 16.9 &       & 19.7 & 2 &125 & 3\\
	 33 & 08 37 31.76&-12 04  04.4 &   55.9 &  -10.2 & 16.8 &       & 17.6 & 2 &127 & 3\\
	 34 & 08 37 31.81&-12 03  31.3 &   56.5 &   22.9 & 18.1 &       & 20.6 & 2 &126 & 1\\
	 35 & 08 37 32.14&-12 04  22.9 &   61.5 &  -28.7 & 18.0 &       & 19.3 & 2 &129 & 1(0.2068)\\
	 36 & 08 37 32.30&-12 03  13.0 &   63.8 &   41.2 & 18.8 &       & 20.4 & 2 &130 & 1(0.1815)\\
	 37 & 08 37 32.39&-12 03  38.4 &   65.1 &   15.8 & 16.2 &       & 17.5 & 2 &132 & 3(0.0828)\\
\hline
3C 215.0 & 09 03 44.16 & 16 58 15.7 &    0.0 &    0.0 & 16.5 & & 18.6 & 3 & 77 & QSO (ref.5 \#1)\\
      1 & 09 03 40.11 & 16 58 38.6 &  -58.0 &   22.9 & 17.6 & & 19.6 & 3 & 20 & 1(ref.9 \#256,0.4125)\\
      2 & 09 03 40.32 & 16 57 15.7 &  -55.1 &  -59.9 & 18.5 & & 20.3 & 3 & 23 & 3\\
      3 & 09 03 40.44 & 16 57 33.2 &  -53.4 &  -42.5 & 19.4 & & 21.7 & 3 & 25 & 2\\
      4 & 09 03 41.48 & 16 57 25.4 &  -38.4 &  -50.3 & 18.1 & & 20.9 & 3 & 38 & 1(ref.9 \#289,0.4093)\\
      5 & 09 03 41.65 & 16 58 03.9 &  -36.0 &  -11.8 & 16.2 & & 18.4 & 3 & 42 & 1(ref.9 \#300,0.2695)\\
      6 & 09 03 42.35 & 16 58 10.0 &  -26.0 &   -5.6 & 17.8 & & 20.1 & 3 & 50 & 1(ref.5 \#17)\\
      7 & 09 03 42.95 & 16 59 03.0 &  -17.3 &   47.3 & 17.8 & & 20.0 & 3 & 60 & 1(ref.9 \#355,0.2682)\\
      8 & 09 03 42.96 & 16 57 07.7 &  -17.2 &  -68.0 & 17.0 & & 18.3 & 3 & 58 & 3\\
      9 & 09 03 43.32 & 16 57 26.6 &  -12.1 &  -49.1 & 17.2 & & 19.6 & 3 & 63 & 1(ref.9 \#362,0.4389)\\
     10 & 09 03 43.64 & 16 57 59.0 &   -7.5 &  -16.7 & 19.1 & & 21.9 & 3 & 66 & 1(ref.5 \#7)\\
     11 & 09 03 43.92 & 16 57 51.8 &   -3.4 &  -23.9 & 17.9 & & 19.9 & 3 & 72 & 1(ref.5 \#6, ref.9 \#390,0.4268)\\
     12 & 09 03 44.21 & 16 58 04.4 &    0.8 &  -11.3 & 19.3 & & 21.3 & 3 & 78 & 3(ref.5 \#5)\\
     13 & 09 03 44.42 & 16 57 10.7 &    3.7 &  -65.0 & 17.1 & & 19.5 & 3 & 80 & 1(ref.9 \#407, 0.2318)\\
     14 & 09 03 44.62 & 16 58 03.6 &    6.5 &  -12.1 & 18.0 & & 20.1 & 3 & 85 & 1(ref.5 \#4, ref.9 \#422,0.4106)\\
     15 & 09 03 44.81 & 16 58 09.4 &    9.4 &   -6.3 & 16.0 & & 17.3 & 3 & 91 & 3(ref.5 \#2)\\
     16 & 09 03 45.33 & 16 57 15.7 &   16.8 &  -60.0 & 18.9 & & 20.9 & 3 & 97 & 3\\
     A  & 09 03 44.47 & 16 58 33.9 &    4.5 &   18.2 & 20.5 & & 22.7 & 3 & 83 & 3(ref.5 \#13)\\
     B  & 09 03 45.04 & 16 58 15.6 &   12.6 &   -0.1 & 20.8 & & 22.5 & 3 & 94 & 1(ref.5 \#11)\\
     C  & 09 03 44.54 & 16 58 12.6 &    5.5 &   -3.1 & 22.3 & &      &   &    & ?\\
\hline
IRAS~09149-6206& 09 14 59.10 & -62  06 54.0  &   0.0  &   0.0 & 11.6&  9.8&&&&QSO\\
 1& 09 14 45.85 & -62  06 54.2  & -92.9  &  -0.2 & 14.6&	   &&&&2\\
 2& 09 14 46.80 & -62  06 48.4  & -86.3  &   5.6 & 15.3&	   &&&&3\\
 3& 09 14 46.92 & -62  05 38.3  & -85.5  &  75.7 & 15.7&	   &&&&3\\
 4& 09 14 47.51 & -62  07 24.8  & -81.3  & -30.7 & 11.6&	   &&&&2\\
 5& 09 14 48.21 & -62  08 15.1  & -76.3  & -81.1 & 16.0&	   &&&&3\\
 6& 09 14 48.76 & -62  07 03.7  & -72.5  &  -9.7 & 11.8&	   &&&&2\\
 7& 09 14 49.54 & -62  07 24.2  & -67.1  & -30.2 & 15.3&	   &&&&3\\
 8& 09 14 49.70 & -62  06 06.3  & -66.0  &  47.7 & 14.8&	   &&&&3\\
 9& 09 14 49.71 & -62  07 31.9  & -65.9  & -37.9 & 15.4&	   &&&&3\\
10& 09 14 49.79 & -62  07 39.0  & -65.3  & -45.0 & 16.4&	   &&&&3\\
11& 09 14 49.80 & -62  07 58.9  & -65.2  & -64.9 & 17.2&	   &&&&3\\
12& 09 14 50.69 & -62  08 02.9  & -59.0  & -68.9 & 16.2&	   &&&&3\\
13& 09 14 50.82 & -62  06 22.4  & -58.1  &  31.6 & 15.4&	   &&&&3\\
14& 09 14 52.99 & -62  05 49.9  & -42.9  &  64.1 & 15.1&	   &&&&3\\
15& 09 14 53.24 & -62  05 32.3  & -41.2  &  81.7 & 16.5&	   &&&&3\\
16& 09 14 53.25 & -62  07 48.8  & -41.0  & -54.8 & 11.8&	   &&&&3\\
17& 09 14 53.33 & -62  06 05.6  & -40.5  &  48.4 & 17.0&	   &&&&3\\
18& 09 14 53.61 & -62  05 25.8  & -38.5  &  88.2 & 16.3&	   &&&&3\\
19& 09 14 53.97 & -62  08 10.7  & -36.0  & -76.7 & 14.5&	   &&&&3\\
20& 09 14 54.26 & -62  05 25.8  & -34.0  &  88.2 & 15.9&	   &&&&3\\
21& 09 14 54.33 & -62  06 12.6  & -33.5  &  41.4 & 15.1& 14.9      &&&&3\\
22& 09 14 54.36 & -62  07 52.4  & -33.2  & -58.4 & 12.4&	   &&&&2\\
23& 09 14 54.46 & -62  05 30.8  & -32.6  &  83.2 & 17.5&	   &&&&3\\
24& 09 14 54.66 & -62  06 07.5  & -31.1  &  46.6 & 16.9&	   &&&&3\\
25& 09 14 54.74 & -62  07 15.9  & -30.6  & -21.9 & 16.9&	   &&&&3\\
26& 09 14 54.86 & -62  07 03.5  & -29.8  &  -9.4 & 17.4& 16.7      &&&&3\\
27& 09 14 55.35 & -62  07 34.3  & -26.3  & -40.3 & 16.6&	   &&&&3\\
28& 09 14 55.58 & -62  06 03.6  & -24.7  &  50.4 & 17.2&	   &&&&3\\
29& 09 14 56.23 & -62  08 08.3  & -20.1  & -74.3 & 16.1&	   &&&&3\\
30& 09 14 56.77 & -62  07 36.3  & -16.3  & -42.3 & 17.5&	   &&&&3\\
31& 09 14 56.81 & -62  07 21.2  & -16.0  & -27.2 & 16.8& 15.4      &&&&3\\
32& 09 14 56.82 & -62  06 49.4  & -16.0  &   4.6 & 17.0& 16.7      &&&&3\\
33& 09 14 56.83 & -62  05 46.2  & -15.9  &  67.9 & 17.0&	   &&&&3\\
34& 09 14 57.19 & -62  07 51.4  & -13.4  & -57.4 & 15.7&	   &&&&3\\
\hline
\end{tabular}
%\end{center}

\end{table*}
\newpage

\begin{table*}
{\bf Table 2.} (Cont.)
%\begin{center}

\begin{tabular}{rrrrrrr|rrrr}
\hline
object & $\alpha$(1950)& $\delta$(1950) & $\Delta \alpha$ & $\Delta \delta$ & $J$ & $K'$ & $R$ & ref & nr. & type($z$) \\
\hline
35& 09 14 57.50 & -62  06 31.8  & -11.3  &  22.2 & 17.4&	   &&&&3\\
36& 09 14 57.62 & -62  05 42.7  & -10.4  &  71.3 & 15.7&	   &&&&3\\
37& 09 14 58.16 & -62  07 44.9  &  -6.6  & -50.9 & 17.3&	   &&&&3\\
38& 09 14 58.38 & -62  05 26.5  &  -5.0  &  87.5 & 13.9&	   &&&&3\\
39& 09 14 58.95 & -62  07 54.1  &  -1.1  & -60.1 & 16.9&	   &&&&3\\
40& 09 14 59.10 & -62  05 49.5  &   0.0  &  64.5 & 17.3&	   &&&&3\\
41& 09 15 00.25 & -62  06 54.2  &   8.1  &  -0.2 & 16.1& 16.3      &&&&3\\
42& 09 15 00.76 & -62  07 37.7  &  11.6  & -43.7 & 16.9&	   &&&&3\\
43& 09 15 01.80 & -62  07 42.5  &  18.9  & -48.5 & 16.7&	   &&&&3\\
44& 09 15 02.16 & -62  07 04.2  &  21.5  & -10.2 & 17.5& 16.8      &&&&3\\
45& 09 15 02.27 & -62  06 54.1  &  22.2  &  -0.1 & 16.0& 16.2      &&&&3\\
46& 09 15 02.39 & -62  06 01.3  &  23.1  &  52.7 & 17.2&	   &&&&3\\
47& 09 15 02.51 & -62  06 43.7  &  23.9  &  10.3 & 18.1&	   &&&&3\\
48& 09 15 02.51 & -62  07 49.0  &  23.9  & -55.0 & 18.2&	   &&&&3\\
49& 09 15 02.66 & -62  07 44.4  &  25.0  & -50.4 & 17.2&	   &&&&3\\
50& 09 15 02.74 & -62  06 35.9  &  25.5  &  18.1 & 14.9& 14.7      &&&&3\\
51& 09 15 02.79 & -62  05 51.9  &  25.9  &  62.1 & 17.0&	   &&&&3\\
52& 09 15 03.26 & -62  05 45.7  &  29.2  &  68.3 & 14.8&	   &&&&3\\
53& 09 15 03.38 & -62  07 08.7  &  30.0  & -14.7 & 17.4& 16.6      &&&&3\\
54& 09 15 03.65 & -62  07 31.8  &  31.9  & -37.8 & 16.5&	   &&&&3\\
55& 09 15 03.74 & -62  06 22.5  &  32.6  &  31.5 & 17.8& 17.2      &&&&3\\
56& 09 15 04.09 & -62  07 11.0  &  35.0  & -17.0 & 17.7&	   &&&&3\\
57& 09 15 04.14 & -62  07 48.8  &  35.4  & -54.8 & 15.9&	   &&&&3\\
58& 09 15 04.43 & -62  05 47.6  &  37.4  &  66.4 & 17.2&	   &&&&3\\
59& 09 15 04.48 & -62  06 03.3  &  37.8  &  50.7 & 17.4&	   &&&&3\\
60& 09 15 04.54 & -62  06 18.6  &  38.2  &  35.4 & 17.8&	   &&&&3\\
61& 09 15 04.81 & -62  07 08.9  &  40.0  & -14.8 & 17.6&	   &&&&3\\
62& 09 15 05.02 & -62  06 02.7  &  41.5  &  51.3 & 16.9&	   &&&&3\\
63& 09 15 05.11 & -62  06 37.3  &  42.2  &  16.7 & 15.5& 15.1      &&&&3\\
64& 09 15 05.49 & -62  07 20.5  &  44.8  & -26.5 & 16.1& 15.6      &&&&3\\
65& 09 15 05.80 & -62  08 04.3  &  47.0  & -70.3 & 16.0&	   &&&&3\\
66& 09 15 05.81 & -62  06 47.0  &  47.0  &   7.0 & 13.8& 13.3      &&&&2\\
67& 09 15 06.03 & -62  07 22.2  &  48.6  & -28.2 & 15.8&	   &&&&3\\
68& 09 15 06.65 & -62  05 42.4  &  53.0  &  71.6 & 17.4&	   &&&&3\\
69& 09 15 06.86 & -62  05 52.2  &  54.4  &  61.8 & 15.4&	   &&&&2\\
70& 09 15 06.99 & -62  05 37.5  &  55.4  &  76.5 & 16.8&	   &&&&3\\
71& 09 15 07.16 & -62  06 29.6  &  56.6  &  24.4 & 13.8&	   &&&&2\\
72& 09 15 07.56 & -62  06 06.8  &  59.4  &  47.2 & 15.7&	   &&&&3\\
73& 09 15 08.02 & -62  06 40.9  &  62.6  &  13.1 & 17.0&	   &&&&3\\
74& 09 15 08.35 & -62  06 54.2  &  64.9  &  -0.2 & 17.0&	   &&&&3\\
75& 09 15 08.44 & -62  07 21.3  &  65.5  & -27.3 & 13.6&	   &&&&1\\
76& 09 15 08.71 & -62  07 48.5  &  67.4  & -54.5 & 16.0&	   &&&&3\\
77& 09 15 09.12 & -62  07 03.5  &  70.2  &  -9.5 & 15.7&	   &&&&3\\
78& 09 15 09.19 & -62  07 24.2  &  70.8  & -30.2 & 14.4&	   &&&&2\\
79& 09 15 09.28 & -62  05 20.8  &  71.5  &  93.2 & 13.4&	   &&&&2\\
\hline
PKS~1011-282 & 10 11 12.20 &-28 16 31.9 &    0.0 &    0.0 & 15.2  &&&&& QSO \\
	   1 & 10 11 03.95 &-28 15 35.4 & -109.0 &   56.5 & 15.8  &&&&&  3\\
	   2 & 10 11 05.01 &-28 16 07.3 &  -95.0 &   24.6 & 18.1  &&&&&  3\\
	   3 & 10 11 07.99 &-28 16 02.3 &  -55.6 &   29.6 & 18.8  &&&&&  3\\
	   4 & 10 11 08.50 &-28 15 25.4 &  -48.8 &   66.5 & 18.2  &&&&&  3\\
	   5 & 10 11 08.54 &-28 16 19.7 &  -48.4 &   12.2 & 16.8  &&&&&  3\\
	   6 & 10 11 09.31 &-28 17 54.3 &  -38.1 &  -82.4 & 18.2  &&&&&  3\\
	   7 & 10 11 09.93 &-28 16 37.5 &  -30.0 &   -5.6 & 17.4  &&&&&  3\\
	   8 & 10 11 09.96 &-28 15 43.0 &  -29.6 &   48.9 & 16.1  &&&&&  3\\
	   9 & 10 11 10.02 &-28 16 07.6 &  -28.8 &   24.3 & 18.8  &&&&&  2\\
	  10 & 10 11 10.26 &-28 16 24.6 &  -25.6 &    7.3 & 16.4  &&&&&  1\\
	  11 & 10 11 10.98 &-28 17 52.2 &  -16.1 &  -80.3 & 18.1  &&&&&  3\\
	  12 & 10 11 11.24 &-28 15 47.8 &  -12.6 &   44.1 & 19.0  &&&&&  3\\
	  13 & 10 11 11.33 &-28 16 11.4 &  -11.5 &   20.5 & 19.5  &&&&&  3\\
	  14 & 10 11 11.91 &-28 16 15.1 &   -3.9 &   16.8 & 17.8  &&&&&  3\\
	  15 & 10 11 12.21 &-28 15 58.2 &    0.2 &   46.1 & 19.1  &&&&&  3\\
	  16 & 10 11 12.22 &-28 15 45.8 &    0.2 &   33.7 & 19.0  &&&&&  3\\
	  17 & 10 11 12.25 &-28 16 09.7 &    0.7 &   22.2 & 19.6  &&&&&  3\\
	  18 & 10 11 12.46 &-28 16 19.7 &    3.5 &   12.2 & 19.2  &&&&&  3\\
	  19 & 10 11 14.12 &-28 16 48.9 &   25.4 &  -17.0 & 17.6  &&&&&  2\\
	  20 & 10 11 14.65 &-28 15 53.4 &   32.4 &   38.5 & 16.8  &&&&&  2\\
	  21 & 10 11 14.67 &-28 17 25.6 &   32.7 &  -53.7 & 18.5  &&&&&  3\\
\hline
\end{tabular}
%%\end{center}
\end{table*}
\newpage

\begin{table*}[h]
{\bf Table 2.} (Cont.)

%\begin{center}
\begin{tabular}{rrrrrrr|rrrr}
\hline
object & $\alpha$(1950) & $\delta$(1950) & $\Delta \alpha$ & $\Delta \delta$ & $J$ & $K'$ & $R$ & ref & nr. & type($z$) \\
\hline
	  22 & 10 11 14.87 &-28 16 48.1 &   35.2 &  -16.2 & 16.3  &&&&&  3\\
	  23 & 10 11 15.19 &-28 15 24.8 &   39.5 &   67.1 & 17.0  &&&&&  3\\
	  24 & 10 11 16.38 &-28 16 40.5 &   55.2 &   -8.6 & 17.9  &&&&&  3\\
\hline
3C 275.1 & 12 41 27.58& 16 39 18.0 &    0.0 &    0.0 & 17.0 & 15.7   & 18.3 & 3 &  32 & QSO \\ 
      1 & 12 41 26.18& 16 39 11.4 &  -20.1 &   -6.6 & 19.1 & 18.1   & 21.8 & 3 &  21 & 1 \\ 
      2 & 12 41 26.49& 16 38 52.9 &  -15.6 &  -25.1 & 18.2 & 16.9   & 20.3 & 3 &  24 & 1(ref.9 \#333,0.4646) \\ 
      3 & 12 41 26.54& 16 40 09.5 &  -14.9 &   51.5 & 18.6 &        & 20.2 & 3 &  25 & 1(ref.9 \#342,0.4664) \\ 
      4 & 12 41 27.19& 16 39 22.4 &   -5.6 &    4.4 & 18.7 & 18.2   & 20.2 & 3 &  26 & 1(ref.9 \#354,0.3374) \\ 
      5 & 12 41 27.46& 16 39 51.2 &   -1.7 &   33.2 & 19.5 &        & 21.4 & 3 &  28 & 1 \\ 
      6 & 12 41 27.76& 16 39 13.3 &    2.7 &   -4.7 & 19.2 & 17.8   & 19.8 & 3 &  34 & 2(ref.9 \#381,0.5570) \\ 
      7 & 12 41 27.94& 16 40 10.6 &    5.2 &   52.6 & 14.7 &        & 15.8 & 3 &  35 & 3 \\ 
      8 & 12 41 28.08& 16 39 02.4 &    7.2 &  -15.6 & 18.9 & 17.9   & 23.1 & 3 &  37 & 1(ref.9 \#393,0.1680) \\ 
      9 & 12 41 28.25& 16 39 17.6 &    9.7 &   -0.4 & 20.5 & 18.2   & 19.8 & 3 &  41 & 3 \\ 
     10 & 12 41 28.64& 16 39 26.4 &   15.3 &    8.4 & 18.1 & 17.1   & 21.9 & 3 &  42 & 1(ref.9 \#425,0.1987) \\ 
     11 & 12 41 29.08& 16 40 06.0 &   21.6 &   48.0 & 16.5 &        & 18.2 & 3 &  48 & 1 \\ 
     12 & 12 41 29.30& 16 39 13.7 &   24.7 &   -4.3 & 19.8 & 18.2   & 21.2 & 3 &  49 & 1 \\ 
     13 & 12 41 29.41& 16 38 50.9 &   26.3 &  -27.1 & 19.2 &        & 21.0 & 3 &  52 & 1(ref.9 \#457,0.4936) \\ 
     14 & 12 41 29.80& 16 38 40.1 &   31.9 &  -37.9 & 18.5 &        & 20.5 & 3 &  56 & 1(ref.9 \#472,0.2566) \\ 
     15 & 12 41 31.70& 16 39 53.2 &   59.3 &   35.2 & 18.7 &        & 20.7 & 3 &  67 & 1(ref.9 \#541,0.4924) \\ 
\hline
PKS 1302-102 & 13 02  55.85&-10 17 16.4 &    0.0 &    0.0 & 13.9 & &15.1 & & & QSO\\
          1 & 13 02  51.50&-10 17 41.6 &  -64.2 &  -25.2 & 14.0 & &15.2 & 6 & 17 & 3\\
          2 & 13 02  51.65&-10 17 20.5 &  -61.9 &   -4.1 & 14.2 & &15.4 & 6 & 19 & 3\\
 	  3 & 13 02  55.01&-10 17 54.4 &  -12.3 &  -38.0 & 17.2 & & 19.0& 6 & 13 & 1\\
 	  4 & 13 02  55.32&-10 18 10.6 &   -7.8 &  -54.2 & 18.8 & & & & & 3\\
 	  5 & 13 02  55.97&-10 16 05.3 &    1.8 &   71.1 & 18.0 & & 20.23& 6 & 1 & 3\\
 	  6 & 13 02  56.38&-10 15 38.8 &    7.8 &   97.6 & 12.1 & & & & & 3\\
 	  7 & 13 02  57.41&-10 17 03.9 &   23.1 &   12.5 & 18.2 & & 20.4& 6 & 5 &1 \\
 	  8 & 13 02  58.11&-10 17 21.8 &   33.4 &   -5.4 & 17.9 & & 19.5& 6 & 8 &1\\
	  9 & 13 03  00.33&-10 16 37.4 &   66.1 &   39.0 & 18.0 & & 19.5 & 6 &4 &3\\
	 10 & 13 03  00.77&-10 16 43.7 &   72.7 &   32.7 & 18.5 & & 21.4 & 6 & 6&3\\
 	 11 & 13 03  01.33&-10 16 52.5 &   80.9 &  -40.7 & 11.1 & & 14.5 & 6 & 9&3\\
	 12 & 13 03  01.33&-10 17 57.1 &   80.9 &   23.9 & 17.9 & & & & & 2\\
\hline
3C 281 & 13 05  22.54& 06 58 14.4 &    0.0 &    0.0 & 16.5 & 15.5 & 17.5 & 4 & 127 & QSO\\
    1 & 13 05  21.19& 06 59 04.2 &  -20.2 &   49.8 & 19.0 &      & 21.7 & 4 & 116 & 1 \\
    2 & 13 05  21.57& 06 58 40.1 &  -14.5 &   25.7 & 19.6 &      & 22.1 & 4 & 119 & 1 \\
    3 & 13 05  21.67& 06 58 08.0 &  -13.0 &   -6.4 & 20.0 &      & 21.3 & 4 & 120 & 1 \\
    4 & 13 05  22.05& 06 58 29.0 &   -7.3 &   14.6 & 18.3 & 17.2 & 21.0 & 4 & 122 & 1 (ref.6 \#17)\\
    5 & 13 05  22.06& 06 57 35.3 &   -7.1 &  -39.1 & 18.7 &      & 21.8 & 4 & 124 & 2 \\
    6 & 13 05  22.08& 06 58 20.1 &   -6.9 &    5.7 & 19.2 & 18.3 & 21.1 & 4 & 123 & 2 \\
    7 & 13 05  22.70& 06 58 35.2 &    2.3 &   20.8 & 18.9 & 17.8 & 21.6 & 4 & 128 & 2 (ref.6 \#1)\\
    8 & 13 05  23.27& 06 57 21.3 &   10.8 &  -53.1 & 19.0 &      &      &   &     &    3\\
    9 & 13 05  23.39& 06 58 06.4 &   12.6 &   -8.0 & 19.7 & 18.4 & 22.2 & 4 & 131 & 3\\
   10 & 13 05  24.01& 06 58 30.9 &   21.9 &   16.5 & 19.4 & 18.5 & 22.1 & 4 & 135 & 1 \\
   11 & 13 05  24.43& 06 58 48.5 &   28.1 &   34.1 & 17.3 & 17.0 & 18.4 & 4 & 138 & 3 (ref.6 \#2)\\
   12 & 13 05  24.99& 06 59 06.5 &   36.5 &   52.1 & 18.5 &      &      &   &     &    1\\
   13 & 13 05  25.31& 06 58 08.1 &   41.2 &  -6.3  & 18.6 & 17.8 & 21.3 & 4 & 139 & 3 (ref.6 \#4)\\
   14 & 13 05  25.32& 06 58 02.4 &   41.4 &  -12.0 & 19.4 & 17.7 & 22.0 & 4 & 140 & 1 \\
\hline
4C~20.33     & 14 22  37.56& 20 13 57.4  &    0.0 &    0.0 & 16.0  & &&&&QSO \\
	   1 & 14 22  33.10& 20 13 08.6  &  -62.8 &  -48.8 & 16.6  & &&&& 3\\
	   2 & 14 22  33.28& 20 14 30.3  &  -60.3 &   32.9 & 18.1  & &&&& 2\\
	   3 & 14 22  35.46& 20 13 24.1  &  -29.5 &  -33.3 & 18.7  & &&&& 2\\
	   4 & 14 22  36.66& 20 13 15.9  &  -12.6 &  -41.5 & 20.1  & &&&& 3\\
	   5 & 14 22  36.67& 20 14 21.2  &  -12.5 &   23.8 & 13.5  & &&&& 2\\
	   6 & 14 22  37.19& 20 14 35.4  &   -5.2 &   38.0 & 18.8  & &&&& 2\\
	   7 & 14 22  37.86& 20 14 07.4  &    4.2 &   10.0 & 19.4  & &&&& 2\\
	   8 & 14 22  38.18& 20 14 56.6  &    8.8 &   59.2 & 16.8  & &&&& 3\\
	   9 & 14 22  38.84& 20 14 37.4  &   18.0 &   40.0 & 18.5  & &&&& 1\\
	  10 & 14 22  38.99& 20 14 33.2  &   20.1 &   35.8 & 18.5  & &&&& 2\\
	  11 & 14 22  39.11& 20 13 32.2  &   21.8 &  -25.2 & 19.5  & &&&& 2\\
	  12 & 14 22  39.97& 20 13 12.4  &   34.0 &  -45.0 & 18.2  & &&&& 1\\
	  13 & 14 22  41.15& 20 14 05.4  &   50.6 &    8.0 & 18.5  & &&&& 3\\
	  14 & 14 22  41.30& 20 13 16.2  &   52.6 &  -41.2 & 18.0  & &&&& 3\\
	  15 & 14 22  41.30& 20 14 20.8  &   52.6 &   23.4 & 19.8  & &&&& 2\\
	  16 & 14 22  41.66& 20 13 12.4  &   57.7 &  -45.0 & 18.0  & &&&& 1\\
	  17 & 14 22  41.93& 20 14 14.8  &   61.5 &   17.4 & 15.1  & &&&& 3\\
\hline
\end{tabular}
%\end{center}

\end{table*}
\newpage

\begin{table*}
{\bf Table 2.} (Cont.)
%\begin{center}

\begin{tabular}{rrrrrrr|rrrr}
\hline
object & $\alpha$(1950)& $\delta$(1950) & $\Delta \alpha$ & $\Delta \delta$ & $J$ & $K'$ & $R$ & ref & nr. & type($z$) \\
\hline
	  18 & 14 22  41.96& 20 15 04.4  &   61.9 &   67.0 & 16.6  & &&&& 3\\
	  19 & 14 22  42.44& 20 14 43.7  &   68.7 &   46.3 & 16.9  & &&&& 3\\
	  20 & 14 22  42.57& 20 13 37.3  &   70.4 &  -20.1 & 17.6  & &&&& 3\\
\hline
4C 11.50 & 15 48  21.20& 11  29 47.0 &    0.0 &    0.0 & 17.4 & 15.7 & 17.9 & 4 & 138 & QSO\\
      1 & 15 48  16.96& 11  29 38.1 &  -62.3 &   -8.9 & 17.2 &      & 19.1 & 4 & 103 & 1\\ 
      2 & 15 48  17.07& 11  29 59.2 &  -60.7 &   12.2 & 18.8 &      & 21.0 & 4 & 104 & 1\\
      3 & 15 48  17.45& 11  29 42.8 &  -55.1 &   -4.2 & 17.3 &      & 19.6 & 4 & 106 & 1\\
      4 & 15 48  19.14& 11  30 19.9 &  -30.3 &   32.9 & 19.4 &      & 21.9 & 4 & 118 & 3\\
      5 & 15 48  19.16& 11  29 24.5 &  -29.9 &  -22.5 & 19.2 &      & 22.6 & 4 & 120 & 1\\
      6 & 15 48  19.19& 11  30 14.4 &  -29.6 &   27.4 & 19.2 &      & 20.2 & 4 & 121 & 1\\
      7 & 15 48  20.53& 11  29 45.1 &   -9.9 &   -1.9 & 16.9 & 16.5 & 19.7 & 4 & 138 & 1(ref.1 \#3, 0.4323)\\ 
      8 & 15 48  21.53& 11  29 46.2 &    4.9 &   -0.8 & 18.1 & 17.4 & 18.8 & 4 & 147 & 3\\
      9 & 15 48  22.78& 11  29 14.5 &   23.3 &  -32.5 & 18.5 & 17.2 & 20.3 & 4 & 154 & 1(ref.1 \#4, 0.4331)\\ 
     10 & 15 48  23.06& 11  30 02.1 &   27.3 &   15.1 & 19.3 &      & 21.1 & 4 & 155 & 1\\
     11 & 15 48  23.26& 11  28 55.1 &   30.3 &  -51.9 & 17.8 &      &      &   &     &  3\\
     12 & 15 48  23.37& 11  30 18.6 &   31.9 &   31.6 & 18.6 & 17.1 & 21.1 & 4 & 157 & 1\\
     13 & 15 48  23.61& 11  29 53.9 &   35.5 &    6.9 & 19.0 &      & 21.4 & 4 & 160 & 3\\
     14 & 15 48  24.88& 11  30 11.3 &   54.1 &   24.3 & 18.4 &      & 19.7 & 4 & 166 & 3\\
      A & 15 48 22.25 & 11 29 45.2  &   15.5 &   -1.8 &      & 19.9 & --   & 4 & 149 & 2 \\
\hline
Mrk~877      & 16 17 56.60 & 17 31 35.0 &    0.0 &    0.0 & 14.2  & &&&&QSO \\
	   1 & 16 17 51.81 & 17 31 48.7 &  -68.5 &   13.7 & 17.1  & &&&& 3\\
	   2 & 16 17 54.17 & 17 30 42.5 &  -34.7 &  -52.5 & 18.7  & &&&& 2\\
	   3 & 16 17 54.26 & 17 32 30.2 &  -33.5 &   55.2 & 19.1  & &&&& 3\\
	   4 & 16 17 54.51 & 17 30 53.5 &  -29.9 &  -41.5 & 17.7  & &&&& 1\\
	   5 & 16 17 54.52 & 17 32 05.6 &  -29.6 &   30.6 & 17.2  & &&&& 3\\
	   6 & 16 17 54.76 & 17 32 39.9 &  -26.3 &   65.0 & 16.4  & &&&& 3\\
	   7 & 16 17 55.12 & 17 31 58.9 &  -21.1 &   23.9 & 19.2  & &&&& 3\\
	   8 & 16 17 56.01 & 17 30 36.2 &   -8.3 &  -58.8 & 17.6  & &&&& 3\\
	   9 & 16 17 56.17 & 17 31 12.3 &   -6.0 &  -22.7 & 16.3  & &&&& 1\\
	  10 & 16 17 57.31 & 17 31 52.0 &   10.2 &   17.0 & 19.2  & &&&& 3\\
	  11 & 16 17 57.55 & 17 31 02.4 &   13.5 &  -32.6 & 19.2  & &&&& 3\\
	  12 & 16 17 57.79 & 17 30 51.2 &   17.0 &  -43.8 & 17.3  & &&&& 2\\
	  13 & 16 17 57.94 & 17 31 30.5 &   19.3 &   -4.5 & 15.8  & &&&& 1\\
	  14 & 16 17 58.19 & 17 30 37.5 &   22.8 &  -57.5 & 15.1  & &&&& 3\\
	  15 & 16 17 58.27 & 17 32 17.3 &   23.8 &   42.3 & 19.8  & &&&& 3\\
	  16 & 16 17 59.37 & 17 30 58.2 &   39.7 &  -36.8 & 17.9  & &&&& 3\\
	  17 & 16 17 59.39 & 17 32 08.3 &   39.8 &   33.3 & 17.6  & &&&& 1\\
	  18 & 16 17 59.62 & 17 32 00.1 &   43.2 &   25.1 & 18.5  & &&&& 3\\
	  19 & 16 17 59.89 & 17 32 11.1 &   47.1 &   36.1 & 18.4  & &&&& 2\\
	  20 & 16 17 59.95 & 17 30 55.3 &   48.0 &  -39.7 & 16.9  & &&&& 3\\
	  21 & 16 18 00.41 & 17 31 32.7 &   54.5 &   -2.3 & 18.8  & &&&& 3\\
	  22 & 16 18 02.28 & 17 31 19.6 &   81.2 &  -15.4 & 18.0  & &&&& 3\\
\hline
3C 334.0 & 16 18 07.40& 17  43 30.5 &    0.0 &    0.0 & 15.7 & & 17.4 & 4 & 128 & QSO(ref.5 \#1)\\
      1 & 16 18 03.83& 17  43 23.4 &  -51.1 &   -7.1 & 20.0 & & 20.9 & 4 & 108 & 3 (ref.6 \#13)\\
      2 & 16 18 04.75& 17  44 21.1 &  -37.8 &   50.6 & 15.9 & &      &   &     &   3\\
      3 & 16 18 05.62& 17  43 03.5 &  -25.4 &  -27.0 & 19.1 & & 21.5 & 4 & 114 &3 (ref.5 \#9)\\
      4 & 16 18 06.50& 17  44 00.5 &  -12.9 &   30.0 & 18.7 & & 21.1 & 4 & 118 & 1 \\
      5 & 16 18 07.23& 17  43 40.3 &   -2.3 &    9.8 & 20.2 & & 22.2 & 4 & 126 & 1 (ref.5 \#19)\\
      6 & 16 18 07.43& 17  43 23.2 &    0.5 &   -7.3 & 19.1 & & 21.6 & 4 & 129 & 1 (ref6 \#10)(ref.5 \#2)\\
      7 & 16 18 07.97& 17  42 58.0 &    8.1 &  -32.5 & 19.1 & & 21.8 & 4 & 133 & 2 (ref6 \#8)(ref.5 \#4)\\
      8 & 16 18 08.57& 17  43 43.5 &   16.7 &   13.0 & 18.6 & & 21.1 & 4 & 138 & 1 (ref6 \#3)(ref.5 \#17)\\
      9 & 16 18 08.79& 17  44 06.2 &   19.9 &   35.7 & 18.8 & & 20.0 & 4 & 140 & 1 \\
     10 & 16 18 09.72& 17  43 16.2 &   33.2 &  -14.3 & 17.6 & & 19.8 & 4 & 144 & 3 (ref.6 \#6)\\
     11 & 16 18 09.84& 17  43 45.5 &   34.8 &   15.0 & 18.5 & & 20.7 & 4 & 146 & 1 (ref.6 \#4)\\
      A & 16 18 08.30& 17 43 17.2  &   12.9 &  -13.3 & 21.6 & & 22.5 & 4 & 135 & 1 (ref.5 \#3)\\
\hline
\end{tabular}
%\end{center}

%\protect\label{autour}
\end{table*}
\newpage

\begin{table*}
{\bf Table 2.} (Cont.)
%\begin{center}

\begin{tabular}{rrrrrrr|rrrr}
\hline
object & $\alpha$(1950)& $\delta$(1950) & $\Delta \alpha$ & $\Delta \delta$ & $J$ & $K'$ & $R$ & ref & nr. & type($z$) \\
\hline
MC~1745+163&17 45 55.75&  16 20 11.5 &    0.0 &    0.0 & 16.2 &15.6&&&&QSO\\
 1&17 45 51.95&  16 19 31.3 &  -54.7 &  -40.2 & 18.7 &	&&&&3\\
 2&17 45 51.98&  16 20 42.9 &  -54.2 &   31.4 & 17.9 &	&&&&3\\
 3&17 45 52.03&  16 19 09.5 &  -53.6 &  -62.0 & 16.3 &	&&&&3\\
 4&17 45 52.42&  16 20 49.4 &  -47.9 &   37.9 & 18.9 &	&&&&3\\
 5&17 45 52.44&  16 19 44.7 &  -47.7 &  -26.8 & 13.3 &13.9&&&&3\\
 6&17 45 52.57&  16 19 15.0 &  -45.8 &  -56.5 & 18.5 &	&&&&3\\
 7&17 45 52.91&  16 20 39.0 &  -41.0 &   27.5 & 19.3 &	&&&&3\\
 8&17 45 52.94&  16 20 15.9 &  -40.4 &    4.4 & 14.7 &15.4&&&&3\\
 9&17 45 52.95&  16 19 35.2 &  -40.4 &  -36.3 & 17.2 &17.1&&&&3\\
10&17 45 53.29&  16 20 44.8 &  -35.4 &   33.3 & 15.7 &16.3&&&&3\\
11&17 45 53.52&  16 19 10.4 &  -32.0 &  -61.1 & 19.2 &	&&&&3\\
12&17 45 53.52&  16 20 15.4 &  -32.0 &    3.9 & 12.9 &13.3&&&&3\\
13&17 45 53.54&  16 19 28.3 &  -31.7 &  -43.2 & 18.9 &	&&&&3\\
14&17 45 54.10&  16 19 13.0 &  -23.8 &  -58.5 & 16.0 &	&&&&3\\
15&17 45 54.22&  16 19 08.9 &  -22.0 &  -62.6 & 16.9 &	&&&&3\\
16&17 45 54.33&  16 20 06.4 &  -20.5 &   -5.1 & 13.6 &14.1&&&&2\\
17&17 45 54.35&  16 20 14.5 &  -20.0 &    3.0 & 18.0 &19.1&&&&3\\
18&17 45 54.36&  16 19 23.3 &  -19.9 &  -48.2 & 17.5 &	&&&&3\\
19&17 45 54.45&  16 19 30.5 &  -18.7 &  -41.0 & 19.2 &	&&&&3\\
20&17 45 54.74&  16 21 22.3 &  -14.6 &   70.8 & 17.5 &	&&&&3\\
21&17 45 54.97&  16 20 23.1 &  -11.3 &   11.6 & 19.0 &	&&&&3\\
22&17 45 55.36&  16 19 23.0 &   -5.5 &  -48.5 & 19.0 &	&&&&3\\
23&17 45 55.37&  16 20 22.9 &   -5.4 &   11.4 & 20.1 &	&&&&3\\
24&17 45 55.38&  16 21 12.4 &   -5.3 &   60.9 & 16.4 &	&&&&3\\
25&17 45 55.47&  16 20 34.7 &   -4.0 &   23.2 & 18.9 &	&&&&3\\
26&17 45 55.65&  16 19 15.0 &   -1.5 &  -56.5 & 16.2 &	&&&&3\\
27&17 45 55.78&  16 19 55.3 &    0.5 &  -16.2 & 17.6 &18.7&&&&3\\
28&17 45 55.80&  16 20 22.9 &    0.8 &   11.4 & 19.8 &	&&&&3\\
29&17 45 55.82&  16 19 18.5 &    1.1 &  -53.0 & 15.9 &	&&&&3\\
30&17 45 56.44&  16 21 17.7 &   10.0 &   66.2 & 18.3 &	&&&&3\\
31&17 45 56.57&  16 20 59.0 &   11.8 &   47.5 & 18.2 &	&&&&3\\
32&17 45 56.90&  16 20 14.7 &   16.5 &    3.2 & 18.7 &	&&&&3\\
33&17 45 57.30&  16 19 50.7 &   22.3 &  -20.8 & 17.4 &17.7&&&&3\\
34&17 45 57.38&  16 20 04.3 &   23.4 &   -7.2 & 16.1 &16.3&&&&3\\
35&17 45 57.69&  16 20 27.0 &   27.9 &   15.5 & 16.5 &17.3&&&&3\\
36&17 45 57.82&  16 19 44.6 &   29.7 &  -26.9 & 18.0 &18.3&&&&3\\
37&17 45 58.04&  16 20 15.2 &   32.9 &    3.7 & 18.9 &	&&&&3\\
38&17 45 58.16&  16 19 26.5 &   34.7 &  -45.0 & 17.1 &	&&&&3\\
39&17 45 58.36&  16 20 25.8 &   37.6 &   14.3 & 17.9 &17.7&&&&3\\
40&17 45 58.54&  16 19 26.8 &   40.3 &  -44.7 & 17.8 &	&&&&3\\
41&17 45 58.56&  16 19 26.7 &   40.5 &  -44.7 & 17.9 &	&&&&3\\
42&17 45 59.11&  16 19 12.1 &   48.5 &  -59.4 & 18.5 &	&&&&3\\
43&17 45 59.30&  16 19 47.0 &   51.1 &  -24.5 & 18.6 &	&&&&3\\
44&17 45 59.45&  16 19 21.4 &   53.3 &  -50.1 & 18.9 &	&&&&3\\
45&17 45 59.89&  16 20 52.8 &   59.6 &   41.3 & 18.9 &	&&&&3\\
46&17 46 00.16&  16 20 08.3 &   63.4 &   -3.2 & 15.7 &	&&&&3\\
47&17 46 00.18&  16 20 51.0 &   63.9 &   39.5 & 17.2 &	&&&&3\\
48&17 46 00.55&  16 21 00.5 &   69.1 &   49.0 & 15.8 &	&&&&3\\
49&17 46 00.56&  16 20 59.9 &   69.2 &   48.4 & 15.7 &	&&&&3\\
50&17 46 00.70&  16 20 36.6 &   71.3 &   25.1 & 17.1 &	&&&&3\\
\hline
4C 11.72 & 22 51 40.55& 11 20 38.7 &    0.0 &    0.0 & 14.4 & 12.7 &15.8$^g$  & 7 &  & QSO\\ 
      1 & 22 51 37.11& 11 20 46.6 &  -50.6 &    7.9 & 18.8 &      &      &   &   &  2\\    
      2 & 22 51 37.12& 11 21 14.7 &  -50.4 &   36.0 & 20.0 &      &      &   &   &  3\\
      3 & 22 51 37.63& 11 20 58.9 &  -42.9 &   20.2 & 18.1 &      &      &   &   &  1\\
      4 & 22 51 38.07& 11 20 56.8 &  -36.5 &   18.1 & 17.7 & 16.6 & 21.1$^g$ & 8  &8   &  1(ref.10 \#2,0.3287)\\
      5 & 22 51 38.64& 11 20 33.4 &  -28.1 &   -5.3 & 17.4 & 16.2 & 20.7$^g$ & 8  &16   &  1(ref.10 \#1,0.3240)\\ 
      6 & 22 51 38.80& 11 20 39.2 &  -25.7 &    0.5 & 18.7 & 18.5 & 20.8~ & 7 & 3 & 1(ref8 \#15)\\ 
      7 & 22 51 39.49& 11 20 29.3 &  -15.6 &   -9.4 & 17.6 & 17.3 & 21.1~ & 7 & 2 & 1(ref8 \#18)\\ 
      8 & 22 51 39.73& 11 19 48.3 &  -12.0 &  -50.4 & 18.1 &      & 21.2$^g$ & 8 & 22 & 1 \\
      9 & 22 51 39.78& 11 20 38.2 &  -11.4 &   -0.5 & 19.4 & 18.1 & 21.4~ & 7 & 1 & 1\\ 
     10 & 22 51 41.66& 11 20 55.8 &   16.3 &   17.1 & 19.4 &      & 21.1$^g$ & 8 & 10   &  1\\
     11 & 22 51 43.77& 11 20 46.1 &   47.4 &    7.4 & 16.1 &      &      &   &   &  2\\
\hline

\end{tabular}
%\end{center}
\begin{footnotesize}

$^g$ {\it g} magnitudes from ref. 8\\
\end{footnotesize}

\end{table*}

\begin{table*}
\caption{Magnitudes of the host galaxies}

\begin{tabular}{lcccccc}
\hline 
\\
object & $J^{host}$ & $J^{host}$/$J^{total}$ & $K'^{host}$ &
$K'^{host}$/$K'^{total}$ & M$_J^{host}$ & M$_{K'}^{host}$ \\ 
& & (\%) & &(\%) & \\   
\\
\hline
\\
A~0401-350A  & 16.0 & 66 & & & -24.8 & \\ 
PKS~0812+020 & 15.8 & 80 & & & -26.5 & \\
PKS~0837-120 & 16.2 & 33 & 13.7 & 76 & -24.4 & -26.9 \\ 
3C~215       & 18.1 & 23 & & & -24.3 &\\ 
IRAS~09149-6206 & 12.2 & 59 & 10.4 & 60 &-25.5 &-27.4 \\ 
PKS~1011-282 & 16.7 & 26 & & & -24.5 & \\ 
3C~275.1     & 17.3 & 80 & & & -25.9 & \\ 
PKS~1302-102 & 14.9 & 40 & & & -26.6 & \\ 
3C~281       & 17.5 & 40 & 15.8 & 73 &-25.8 &-27.5 \\ 
4C~20.33     & 16.3 & 73 & & & -28.1 & \\
4C~11.50     & 17.8 & 71 & & & -24.7 & \\ 
Mrk 877      & 15.0 & 50 & & & -24.3 & \\ 
3C~334       & 16.3 & 57 & & & -26.8 & \\ 
MC~1745+163  & 17.9 & 21 & 16.2 & 58 & -24.3 & -26.2 \\ 
4C~11.72     & 15.6 & 34 & & & -26.2 & \\ 
\\
\hline
\end{tabular}

\begin{footnotesize}
To calculate absolute magnitudes we have used H$_0$= 50
km~s$^{-1}$~Mpc$^{-1}$ and q$_0$=0\\
\end{footnotesize}

\protect\label{tabhost}
\end{table*}

\end{document}